%% file: main.tex
\documentclass{article}

\usepackage{PRIMEarxiv}

\usepackage{graphicx}
\usepackage{authblk}
\usepackage{amsmath}
\usepackage{xcolor}
\usepackage{makecell}
\usepackage{bbding}
\usepackage{soul}
\usepackage{mhchem}
\usepackage{siunitx}
\usepackage{lineno}
\usepackage{hyperref}

\newcommand{\spectrofunc}{f}

\newcommand{\inputvar}{x}
\newcommand{\newinputvar}{x^*}
\newcommand{\newmeas}{f^*}
\newcommand{\newobs}{y^*}
\newcommand{\numpastmeas}{n}
\newcommand{\pastlocs}{\mathbf{x}}
\newcommand{\pastmeas}{\mathbf{f}}
\newcommand{\pastobs}{\mathbf{y}}
\newcommand{\covar}{\mathbf{K}}
\newcommand{\mean}{\boldsymbol{\mu}}
\newcommand{\traintestcovar}{\boldsymbol{\kappa}}
\newcommand{\noisecovar}{\mathbf{N}}
\newcommand{\posteriormean}{m}
\newcommand{\posteriorvariance}{\sigma^2}
\newcommand{\posteriorstd}{\sigma}
\newcommand{\kernelhp}{\boldsymbol{\theta}}
\newcommand{\kernelfunc}{k_{\kernelhp}}
\newcommand{\deepkernel}{g_\mathbf{\Phi}}
\newcommand{\matern}{Mat\'{e}rn}
\newcommand{\acqf}{a}
\newcommand{\acqfg}{a_g}
\newcommand{\acqfr}{a_r}
\newcommand{\acqfweightg}{\phi_g}
\newcommand{\acqfweightgg}{\phi_{g'}}
\newcommand{\acqfweightr}{\phi_r}
\newcommand{\fittedspectrum}{\tilde{f}}
\newcommand{\edgeloc}{\inputvar_e}
\newcommand{\weightfloor}{w_{s,\text{floor}}}
\newcommand{\degc}{$^\circ$C}
\newcommand{\insitu}{\emph{in-situ}}

\DeclareMathOperator*{\argmax}{argmax}

\title{Demonstration of an AI-driven workflow for dynamic x-ray spectroscopy}

\author{
  \textbf{Ming Du$^*$, Mark Wolfman, Chengjun Sun, Shelly D. Kelly, Mathew J. Cherukara} \\
  Advanced Photon Source\\
  Argonne National Laboratory\\
  Lemont, IL 60439, USA \\
  $^*$\texttt{mingdu@anl.gov}
}

\begin{document}

\include{government_license}
\newpage

\maketitle

\begin{abstract}
X-ray absorption near edge structure (XANES) spectroscopy is a powerful technique for characterizing the chemical state and symmetry of individual elements within materials, but requires collecting data at many energy points which can be time-consuming. While adaptive sampling methods exist for efficiently collecting spectroscopic data, they often lack domain-specific knowledge about XANES spectra structure. Here we demonstrate a knowledge-injected Bayesian optimization approach for adaptive XANES data collection that incorporates understanding of spectral features like absorption edges and pre-edge peaks. We show this method accurately reconstructs the absorption edge of XANES spectra using only 15-20\% of the measurement points typically needed for conventional sampling, while maintaining the ability to determine the x-ray energy of the sharp peak after absorption edge with errors less than 0.03 eV, the absorption edge with errors less than 0.1 eV; and overall root-mean-square errors less than 0.005 compared to compared to traditionally sampled spectra. Our experiments on battery materials and catalysts demonstrate the method's effectiveness for both static and dynamic XANES measurements, improving data collection efficiency and enabling better time resolution for tracking chemical changes. This approach advances the degree of automation in XANES experiments reducing the common errors of under- or over-sampling points in near the absorption edge and enabling dynamic experiments that require high temporal resolution or limited measurement time.
\end{abstract}


\section{Introduction}

X-ray absorption spectroscopy measures the probability of materials to absorb x-rays impinging on them with different x-ray energies. These x-ray energy-space measurements are correlated with the atomic or molecular properties of materials, which makes spectroscopy an important tool for material characterization. In particular, X-ray absorption near-edge structure (XANES) spectroscopy is an indispensable technique in revealing
chemical properties such as the oxidation and bonding states of specific atoms within the material, and has been used in studying actinide bonding \cite{silva_nature_comm_2024}, spatial phase mapping \cite{ade_science_1992, wan_science_2024}, and investigating the charging/discharging cycles of batteries \cite{yang_natsustain_2024}, to name a few. When x-rays impinge on the atoms of a material, they are
absorbed significantly more when the x-ray energies surpass the binding energy of core electrons of a specific atom type or element in the material resulting in an absorption discontinuity called an absorption edge. Hence, each element exhibits multiple absorption edges representing the binding energies of the elements' electrons. The sharp rise in the absorption coefficients are called the $K$-edge (for $1s$ electrons) and $L_1$, $L_2$, and $L_3$-edges (for $2s$ and $2p_{1/2}$ and $2p_{3/2}$ electrons). The absorption spectrum of an element in materials are altered by factors such as the oxidation and bonding symmetry of the element. Since the absorption edge is a direct measure of the x-ray energy required to excite a core electron to unoccupied electronic states, the absorption edge shifts depending on the oxidation state of the element, and pre-edge peaks appear when transitions are allowed to unoccupied electronic states below the Fermi level (unoccupied bound states).  Open states at the Fermi level result in strong absorption at or just above the absorption edge itself which is called a ``white line''. Fine structure oscillations beyond the absorption edge are due to the interference of the excited photo electron with the electrons of the neighboring atoms. Long multiple scattering photo-electron paths are possible for low energy photo-electrons which can give rise to resonant features just above the absorption edge.  \cite{zabinsky_prb_1995, jacobsen_xraymicroscopy_2019}. Measuring these fine features therefore reveals the phase and chemical state of elements within the sample.

XANES spectroscopy is traditionally performed by sequentially scanning
the x-ray energy across the absorption edge using an energy-adjustable monochromator (such as a double crystal monochromator). During the scan the spectra are recorded by collecting the incident and transmitted x-ray intensity for each x-ray energy point. This is called a transmission mode measurement. The absorption coefficient is obtained by taking the log ratio of the incident to transmitted x-ray intensity following Beer's law. The absorption coefficient is also proportional to the fluorescent x-rays that are produced as a result of the core-hole produced by exciting the core-electron. Such that the absorption coefficient can also be obtained by measuring the ratio of fluorescent x-ray intensity normalized by the incident x-ray intensity called fluorescence-mode measurement.  The time consumed by the measurement is thus proportional to the number of energy points scanned and the time spent at each of those points to collect sufficient measurement statistics. There are two commmon ways in which the monochromator is scanned in an XANES measurement. Historically, the monochomator is stepped from one energy point to the next energy point. The step size in x-ray energy between the points is varied depending on the part of the x-ray spectrum being measured. Typical step sizes are 5 to 10 eV in the smoothly varying pre-edge region, 0.1 to 1 eV in the fast changing absorption edge region, and 1 to 5 eV in the more slowly varying post edge region. One common failure mode of this style of data collection is the improper definition of the boundaries of each region. The 5 eV step size used for the pre-edge region can cause a pre-edge feature to be completely missed; errors also occur if the step size in the absorption edge region is too small (the monochromator doesn't actually move) or too large (features are not well defined) \cite{meyer_jcatalysis_2024}. More recently, monochromators have become capable of slew scanning and a uniform grid of energy points becomes the simplest data collection mode. This results in over-sampling in some regions (pre- and post-edge) of the spectra necessitating post processing such as rebinning of data points. In either collection mode, if the scan is set up properly, the density of measured points are restricted to the highest point density required for a rather large region of the entire spectrum, which often results in more points than what
is needed to sample the essential information. 

The number of points to sample is a even greater concern for dynamic XANES experiments.
In a dynamic process where the sample's chemical properties undergoes continuous variation often under external stimulii such as voltage \cite{yang_natsustain_2024} and temperature \cite{pascarelli_natrev_2023}, XANES
can be done repeatedly during the process to characterize the sample' states at different time
points. For example, 
for a chemical process that involves changes in the oxidation state of a certain element, 
the absorption edge's position and the fine structures near it in its XANES spectra before
and after the process often differ, and
measuring XANES spectra during the transition thus allows one to find out the progress of the process at the time when that spectrum is measured. In turn, plotting the transition progress against
time reveals the kinetics of the transition. The time resolution of dynamic XANES is almost
as important as its energy resolution in each spectrum. A sampling strategy that shortens
the per-spectrum acquisition time by reducing the number of points while maintaining the energy
resolution of the spectra is thus highly desired.

Adaptive sampling strategies have been developed to break the stalemate between resolution 
and experiment time. Adaptive sampling algorithms run during an experiment, receive the measured
values or instrument readouts in real-time, analyze the updated data, and suggest the points
to measure or actions to take in order to maximize the information gain from the measured
object. The points measured under the guidance of adaptive sampling algorithms are often sparse
and do not necessarily lie on a regular grid. Given the measured values, one may reconstruct
the measured signal in an arbitrarily dense grid through interpolation assuming that the missing data lies smoothly between the measured points. A successful adaptive sampling algorithm
is distinguished by its ability to suggest a small number of measurement points and yield a spectrum with maximum information. 

A survey of existing adaptive sampling methods leads us to categorize them into data-driven approaches
and Bayesian optimization approaches. Data-driven approaches refer to those that use a trained model
to predict the best action to take based on measurements made in the past and the current
environment (such as sample conditions or probe positions). One representative technique
is reinforcement learning \cite{betterton_icra_2020}, where a model is gradually trained to
make the best move (\emph{e.g.}, to suggest the next location to sample) using the reward
(\emph{e.g.}, the reduction of the reconstructed signal's error after taking a certain measurement) 
from past actions. 
Another variant is represented by algorithms named SLADS-Net \cite{zhang_electronicimaging_2018}
and FaST \cite{kandel_natcomm_2023}. Used in scanning microscopy to sample a 2D image, 
these methods employ a neural network trained to predict the reduction of the reconstructed
image's error after measuring at a point, with the positional local contextual information
of that point as the input to the network. During an experiment, the error reduction is predicted
on a set of points, and the ones with the highest predicted error reduction are sampled in
the next step. While they have demonstrated success in reported use cases, a common concern about
data-driven approaches, as noted in \cite{noack_natrevphys_2021}, 
is that they need to be trained on data from previous experiments, but the transferrability
of the trained models to test samples is unclear due to the lack of interpretability. 

Bayesian optimization (BO) approaches on the other hand use a random process, which is
often a Gaussian process (GP), to model
the expected values and the uncertainties of the signal being measured. 
GP-based BO has been used for the adaptive sampling in various
characterization techniques. Noack \emph{et al.} \cite{noack_natrevphys_2021} introduced the applications of such
a Bayesian optimization method in the 2D space-domain sampling of spatially resolved 
x-ray scattering (also reported in \cite{noack_scirep_2019}), 
Fourier transform Infrared absorption spectroscopy (FTIR) (also reported in \cite{holman_commbio_2020}), 
angle-resolved photoemission spectroscopy (ARPES), and the reciprocal space-energy-domain sampling of
inelastic neutron scattering. In 
1D spectroscopy where sampling is done in the energy domain,
Bayesian optimization has also been used for x-ray magnetic circular dichroism spectroscopy 
\cite{ueno_npjcompmat_2018} and x-ray absorption spectroscopy (XAS) \cite{zhang_npjcompmat_2023}. 
Closer to our intended application is \cite{cakir_mlscitech_2024},
where Bayesian optimization is applied for grazing-exit XANES. 

\subsection{Gaussian process and Bayesian optimization}

For a GP, uncertainty quantification
is realized through the kernel function $\kernelfunc(\inputvar_1, \inputvar_2)$ that estimates the
correlation between two points given their positions in the measured space, $\inputvar_1$ and $\inputvar_2$.
Commonly, a stationary kernel function such as the radial basis function or 
\matern~kernel \cite{matern_spatialvariation_1986}
is used, which estimates the correlation purely by their distance, \emph{i.e.},
$\kernelfunc(\inputvar_1, \inputvar_2) = \kernelfunc(|\inputvar_1 - \inputvar_2|)$, where $\theta$ is the set of hyperparameters defining the kernel's behavior.
If point 1 is a measured point and point 2 is unmeasured, a larger $|\inputvar_1 - \inputvar_2|$
generally results in a smaller correlation and thus higher uncertainty for the value of point 2,
and the decay rate of correlation is set by the lengthscale of the kernel which is a part
of $\kernelhp$ and can be estimated using a sparse collection of initial measurements.
A GP still needs to be constructed with data, but the amount of data needed is far 
less than data-driven approaches
and they can come from a sparse set of initial measurements at the beginning of the current
experiment. Once the GP is constructed, 
it then yields the estimated value and uncertainty for any point $\inputvar$ in the sampled space. 
In statistics terms, these are the posterior mean $\posteriormean(\inputvar)$
and posterior standard deviation $\posteriorstd(\inputvar)$. These quantities are used to
formulate an acquisition function $\acqf(\inputvar)$ that quantifies the benefit of
taking a measurement at $\inputvar$. The next measurement is done at the maximum point of 
$\acqf(\inputvar)$, \emph{i.e.}, $\newinputvar = \argmax_\inputvar \acqf(\inputvar)$.
The measured value $\newobs$ along with $\newinputvar$ is used to update the GP,
which then yields a new suggestion. This process is repeated until a stopping criterion is met.

Compared to purely data-driven approaches, BO has the advantages of not requiring a large amount of training data, and is more resistant to drift from training set distribution. Moreover, BO provides uncertainty quantification naturally, and it is straightforward to define and regulate the behavior of a BO algorithm analytical using physical and statistical knowledge. These features make BO a favorable candidate for adaptive sampling. 

\subsection{Domain knowledge-aware Bayesian optimization}

From the perspective of adaptive sampling algorithm design, a unique characteristic of spectroscopic techniques like XANES is that we often have good
prior knowledge about the spectra being measured. XANES spectra are typically composed
of a pre-edge region, pre-edge peaks a few eVs below the edge, the absorption edge,
and the fine structure above the edge. We roughly know the forms they take: the pre-edge region should have a smooth slope, the absorption edge is characterized by a large
discontinuity, while the post-edge features are wavy undulations. Injecting this knowledge to a adaptive
sampling algorithm and instructing it to sample more densely around the varying regions 
can greatly enhance the efficiency of sampling as it reduces the unnecessary exploration in 
the less interesting slowly varying regions. 

Moreover, a XANES spectrum in a dynamic process can be well-approximated by the linear combination of at least 2 spectra collected at different times of the process \cite{kelly_2015, calvin_2024}. It follows that if the spectrum reconstructed during an experiment is fitted with reference spectra, higher fitting residue should indicate regions that need more measurements. 

However, GP-BO with standard
components do not automatically take advantage of these knowledge and properties of XANES. 
In \cite{ueno_npjcompmat_2018}, the use of an
acquisition function only containing the posterior variance makes the algorithm unaware of the 
spectrum's intensity as the posterior variance is only dependent on the positions (energies)
but not the observed values of measurements (see Eq.~\ref{eq:posterior_variance}).
\cite{cakir_mlscitech_2024} uses the upper confidence bound (UCB) of 
$\posteriormean(\inputvar) + \kappa\posteriorstd(\inputvar)$ as the acquisition function, but
it can be argued that the added posterior mean term merely makes the algorithm tend more to
sample in regions where the spectrum intensity is high, which is not necessarily desirable 
since some features (such as the pre-edge peak) that are low in intensity might be of higher
importance compared to high-value yet featureless regions (such as a smoothly varying post-edge region).
We are thus motivated to develop a method that is injected with this structural knowledge and leverages it to optimize the sampling distribution in the energy domain.

How can the structural and data-derived knowledge be incorporated in a GP-BO algorithm? \cite{noack_natrevphys_2021} listed 3 strategies for domain knowledge injection, 
namely (1) using a (potentially non-stationary) kernel function that better reflects the 
lengthscale of signal variations, (2) constraining the fitting of kernel hyperparameters, 
and (3) designing an acquisition function that better reflects the benefits and costs
of measuring at given locations. We consider (3) as the most straightforward approach because
the maximum of the acquisition function is what directly dictates the suggestion of a BO-based
algorithm. Also, since the acquisition function is defined in the same domain as the
input variable of the measured signal (energy in the case of XANES), we can easily apply
our knowledge about the energy-domain structure of XANES spectra. A number of previous works
have adopted the acquisition function engineering strategy for other characterization techniques: in an adaptive sampling algorithm designed for x-ray scattering \cite{noack_scirep_2020},
an acquisition function is designed to take into account the posterior uncertainty, 
the value and spatial gradient of the estimated (\emph{i.e.}, densely reconstructed) signal, and the cost of taking a particular
measurement (\emph{e.g.}, the time consumed by moving the motor). In \cite{zhang_npjcompmat_2023},
the authors use a GP to model the difference between the XAS spectrum predicted 
with a physics model and the observed values, and the acquisition function for sampling is the UCB
of this difference; the hyperparameters used in the physics model are searched by another
BO. It should be noted that an acquisition function designed specifically for a certain 
characterization technique is not guaranteed to be applicable to another technique due to
the difference in the prior knowledge and assumptions involved.

Here we introduce our adaptive sampling algorithm where an acquisition function is designed
and engineered with the prior structural knowledge about XANES spectra. 
As described in section \ref{sec:methods}, the main design features of our acquisition
function include:
\begin{itemize}
    \item A first- and second-order gradient component that returns higher values at peaks,
    values and slopes of the reconstructed spectrum;
    \item A fitting residue component that responds to the difference between
    the reconstructed spectrum and the spectrum linearly fitted using 2 reference spectra is high;
    \item An acquisition reweighting function that automatically detects the location and width
    of the absorption edge, based on which it reduces acquisition function values 
    in the pre-edge smoothly varying region and boosts its values in the high-variation region above the edge.
\end{itemize}

\begin{figure}
    \centering
    \includegraphics[width=1\linewidth]{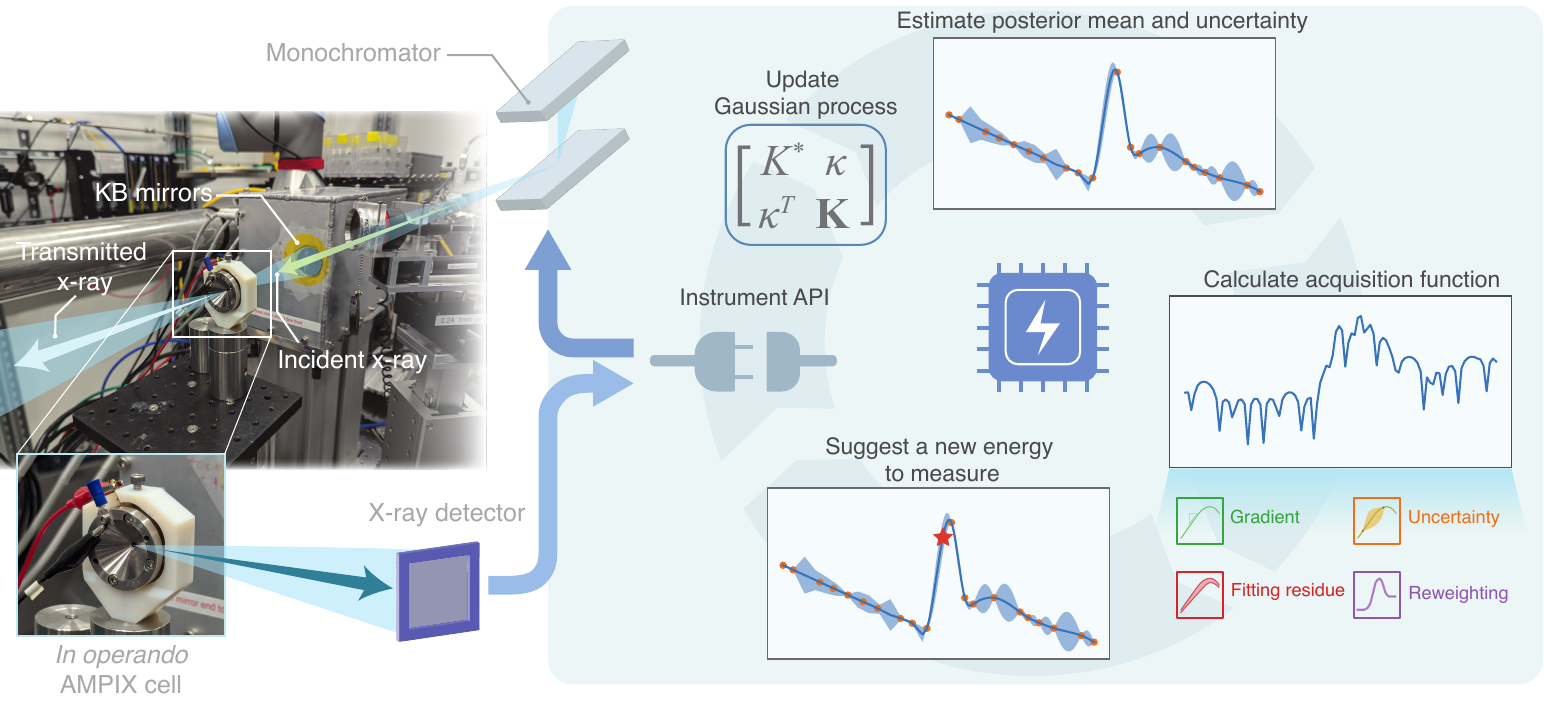}
    \caption{Diagram showing the workflow of our adaptive sampling method, illustrated using the real-world battery cell experiment demonstrated in this paper. The backend algorithm receives data from the x-ray detector and sends data to the monochromator through the instrument API powered by Bluesky \cite{allan_synchrotron_rad_news_2019}. With each measured energy, the algorithm updates its internal state, computes the comprehensive acquisition function and directs the monochromator to measure at a new energy. This cycle is repeated until the stopping conditions are met.}
    \label{fig:diagram}
\end{figure}

We test our algorithm in three simulated case studies and one real-world
experiment. 
We first assess our algorithm on the sampling of single XANES spectra and closely examine the behavior
of the algorithm during data collection. 
Comparing the reconstructed spectra with the densely sampled ground truths, we show that the overall root-mean-squared (RMS) errors of the normalized spectra are generally below 0.005; moreover, in most cases, our method requires only 15--20\%
of the number of points measured in the ground truth. The rate of error reduction using our method is faster both than sampling
on a uniform grid and a BO algorithm that uses an uncertainty-only acquisition function. 
We then use our algorithm to collect spectra in dynamic XANES experiments tracking the 
evolution of the chemical states of certain elements in materials during dynamic processes. The percentage of transition of the chemical state is calculated
for each spectrum during the process through reference spectrum fitting, and the curve of
the transition progress is compared with that obtained using traditionally densely sampled spectra
to show that our method reveals information about the kinetics of the process as
accurately as traditional sampling while requiring a much smaller number of measurements. Additionally, we show that the adaptively sampled spectra allow one to determine the energy of the maxima of the white line with an error of at most 0.03 eV, and the absorption edge with an error of at most 0.1 eV.  
Finally, we also demonstrate the use of our algorithm in a live experiment
at a synchrotron beamline where it communicates with beamline instruments
in real-time to guide the \insitu~XANES measurements of a battery 
electrode. An illustration of our method's workflow using this experiment is shown in Fig.~\ref{fig:diagram}.

\section{Results}

\subsection{Simulated studies}

\subsubsection{Sampling of a single spectrum from the YBCO sample}

\begin{figure}
    \centering
    \includegraphics[width=1\linewidth]{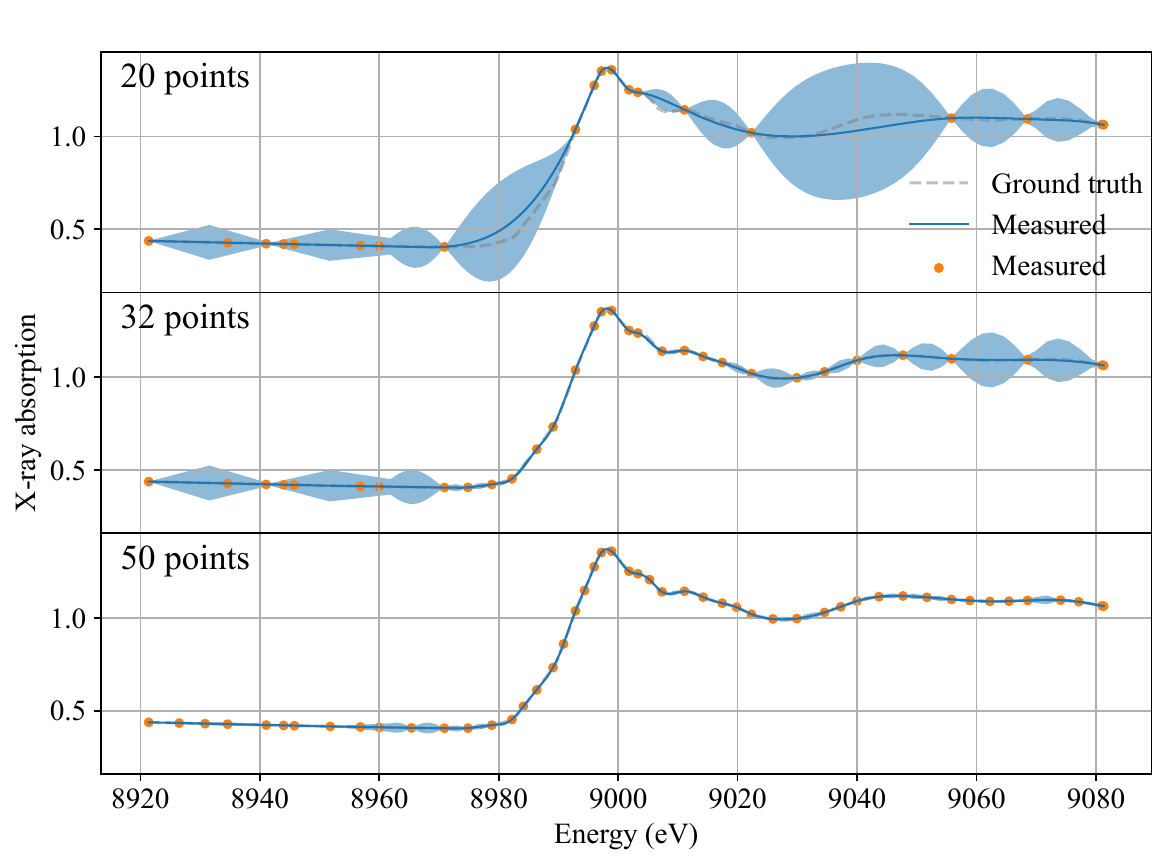}
    \caption{Intermediate reconstructed spectra, posterior standard deviation, 
             measured data points, and true spectrum for the YBCO data. Data are plotted
             without normalization.
             The posterior standard deviation at each point is represented by half of the vertical
             length of the shaded area.}
    \label{fig:ybco_intermediate}
\end{figure}

We first apply our method in sampling the XANES spectrum of yttrium barium copper oxide (YBCO), 
a compound exhibiting high-temperature superconductivity. 
The data for this simulated test case was collected in a previous
experiment with manually chosen scan grid; in the range between
8920 and 9080 eV,
the original spectrum was acquired with a step size of 10 eV below
8960 eV, 0.3 eV between 8960 eV and 9012 eV, and 1.5 eV above 9012 eV. This constitute a total 
of 218 points in the ground truth data. For our test,
we get the virtually measured values at the energies 
suggested by the algorithm through linearly interpolating the ground truth. 
The data used for sampling are unprocessed, meaning their values may contain sample-specific
scaling in edge height, and a slope in the energy domain.
This simulates the data condition that our sampling algorithm will encounter in an actual
experiment.
As mentioned in section \ref{sec:early_stopping}, the algorithm is allowed to stop before the set number of measurements is reached if the maximum weighted uncertainty drops below a threshold. 
The maximum posterior standard deviation threshold for early stopping was set to $t = 0.03$.
We used 20 randomly positioned initial measurements to build the GP and fit the lengthscale parameter;
the latter was found to be 10.2 eV. 
The stopping criterion was triggered after 30 additional measurements giving a total
of 50 measurements. The reconstructed spectrum given by the spline interpolation
of the measurements, the posterior standard deviation (represented by half of the vertical length
of the shaded area at each point), the measured points, and the ground truth are shown in 
Fig.~\ref{fig:ybco_intermediate}. These quantities are plotted for 20, 32, and 50 measured points. . 
With the constraint of value-aware terms in the comprehensive acquisition function and the
acquisition reweighting function, the algorithm barely sampled points in the pre-edge smoothly varying region
below 8960 eV during the first 10 -- 15 guided measurements. 
Rather, it allocated most of the measurements on the absorption edge
at 8080 -- 9000 eV and the post-edge features above 9000 eV. In particular, the fast
post-edge undulations between 9000 and 9030 eV are most densely sampled in the first 10 -- 15
guided measurements. The algorithm started to explore the pre-edge and post-edge smoothly varying regions
when the total number of measurements went beyond 32, at which point uncertainty became a
more important driving force.

To quantitatively evaluate the accuracy of the spectrum sampled 
with our algorithm, we calculate
the RMS error between the reconstructed spectrum and the ground truth after each
new point is measured. For result interpretability, 
both spectra are normalized for RMS error calculation, such that the linear slope background
is subtracted from the whole spectra, and the curvature in the post-edge region is corrected.
This is done following the approach documented for the Athena package \cite{ravel_intltblcrystal_2020}:
first, a straight line $\spectrofunc = F_1(\inputvar)$ is fitted using pre-edge data 
and a quadratic function $\spectrofunc = F_2(\inputvar)$ is fitted using
post-edge data. The fitting ranges are chosen manually. The fitted line is extrapolated to the 
entire energy range and subtracted from the spectrum. The difference between the two fitted
functions at the edge energy $\Delta\spectrofunc = F_2(\inputvar_0) - F_1(\inputvar_0)$ 
is estimated to be the edge height, and the spectrum is scaled by $1 / \Delta\spectrofunc$.
 Subsequently, the spectrum is flattened by subtracting
$[F_2(\inputvar) - F_1(\inputvar) - \Delta\spectrofunc] / \Delta\spectrofunc$ for
$\inputvar > \inputvar_0$. This completes the normalization, after which the edge height becomes 1 and the pre-edge region lies on zero and the post edge oscillates about one. We use the full raw spectra that includes the data outside the
adaptively sampled range ($<8920$ eV and $>9080$ eV) to fit $F_1$ and $F_2$ because they 
contain more smoothly varying segments. The other important feature of the XANES data is the absorption edge position in x-ray energy. This is often defined as the maximum of the first derivative of the absorption edge or the first zero crossing of the second derivative.

\begin{figure}
    \centering
    \includegraphics[width=1\linewidth]{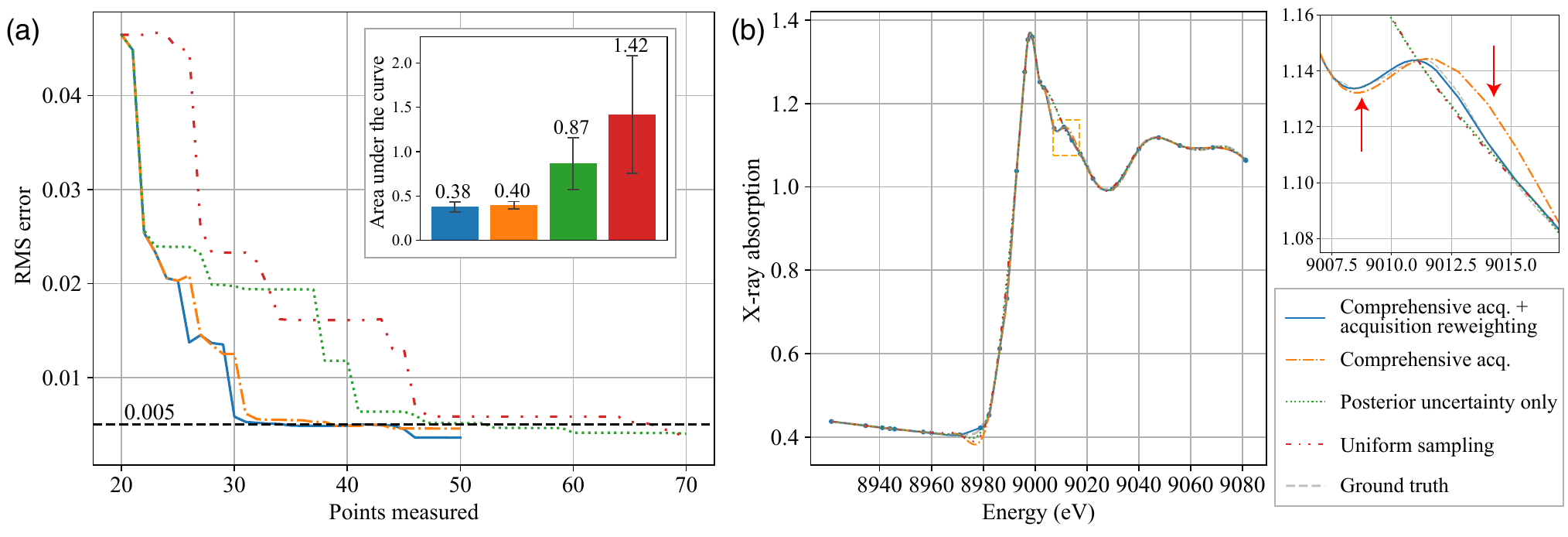}
    \caption{Results of single-spectrum sampling of the YBCO sample. Legend at the bottom right corner applies to both sub-figures.
             (a) Convergence of RMS error with our comprehensive acquisition function and acquisition reweighting,
             and its comparison with cases (\emph{i}) with comprehensive acquisition but without reweighting, 
             (\emph{ii}) with posterior uncertainty-only acquisition
             function, and (\emph{iii}) with uniform sampling. 
             The inset bar chart on the side shows the areas under 
             the curve before the 50th measurement for each case with
             the same color coding to the RMS error curves. 
             The heights of the bars and the numbers indicate the averages of the 
             AUCs over 5 repeated runs with different initial points, and the error bars
             represent their standard deviations.
             (b) Reconstructed spectra using these methods after
             30 points are sampled (initial measurements included). The inset shows the magnified spectra
             within the energy range between 9007 and 9017 eV. This offers a direct comparison between
             the comprehensive acquisition method with and without reweighting, with the latter exhibiting
             larger errors in the plotted region.}
    \label{fig:ybco_comparison}
\end{figure}

We compare the RMS values of our method with three reference cases ablated from our
method: (1) a GP-BO algorithm that is otherwise the same as our proposed method but without the 
acquisition reweighting; (2) a GP-BO algorithm that only uses posterior standard
deviation as the acquisition function; and (3) uniform sampling. The uniform sampling scheme
progressively bisects between measured points: assuming the measured range is normalized to [0, 1],
the first 2 points to measure after the initial measurements are 0 and 1, and subsequently it
measures following the sequence of $[0.5, 0.25, 0.75, 0.125, 0.375, 0.625, 0.875, ...]$. 
This ensures that the sampled points have a quasi-even coverage of the whole spectrum all the
time, while their density gradually increases. Fig.~\ref{fig:ybco_comparison}(a) indicates
that the RMS error of uniform sampling is almost always higher than other techniques, and
takes more measurements (about 65 points) 
to drop below 0.005 RMS error. GP with posterior uncertainty-only acquisition function
reaches 0.005 RMS error after about 46 measurements. In contrast, when the comprehensive acquisition
function is used, 0.005 RMS error is reached after 38 measurements. When acquisition
function reweighting is enabled, 0.005 RMS error is reached with only 34 measurements,
or 15.6\% of the ground truth points. 
An alternative metric to compare the error reduction speed of these methods is the area under
the curve (AUC) of the RMS error convergence plots. This method is also employed in
\cite{godaliyadda_ieeetci_2018}. Smaller AUCs indicate faster error reduction. For each method, we ran the spectrum sampling 5 times and plotted the average AUCs before the 50th measurement in the inset of
Fig.~\ref{fig:ybco_comparison}. The error bars indicate the standard deviations over the 5 runs with different initial measurements. 
The AUCs of methods with the comprehensive acquisition function are obviously smaller than 
those of uncertainty-only BO and uniform sampling. Comprehensive acquisition function
with reweighting has the smallest average AUC of 0.38 ($\pm $0.06), which marks the best performance 
among the tested methods. Additionally, the AUCs of the methods with comprehensive acquisition function is more stable and less sensitive to initial measurements as indicated by their smaller standard deviations. We do note that the AUC advantage of the case with acquisition function reweighting to the case without reweighting is not drastically large, because the former may have slightly
higher error in the pre-edge smoothly varying region due to sparser sampling there. 
However, the merit of reweighting is more pronounced in the feature-rich post-edge region. 
Fig.~\ref{fig:ybco_comparison}(b) shows reconstructed spectra (during the first run) using the 4
listed methods after the 30th point is measured along with the ground truth. 
Uniform sampling and GP-BO with uncertainty-only acquisition function have not yet sampled
critical inflection points like the one at 9008 eV, resulting in large deviations
from the ground truth in the post-edge features. The cases using the comprehensive acquisition
function with and without reweighting are similar, but closer look at the region between
9007 and 9017 eV (outset of Fig.~\ref{fig:ybco_comparison}(b)) reveals better accuracy given by the reweighting-enabled algorithm
due to more samples measured in this fast varying area. The RMS error within this local range is 0.001 for comprehensive acquisition function with reweighting, and 0.005 without.

\subsubsection{Sampling in a dynamic XANES experiment on the LTO sample}

In this section, we apply our adaptive sampling method to a dynamic Ti $K$-edge XANES experiment intended for tracking
the progress and studying the kinetics of the phase transition of lithium titanium oxide (LTO), a battery
material. Raw XANES data were collected during the phase transition of the material at 2 different temperatures,
50\degc~and 70\degc. 
In the range between 4936 and 5006 eV, the ground truth was sampled with a uniform
step size of 0.5 eV, resulting in 141 points per spectrum. 
128 spectra were collected for the phase transition at 50\degc, while 14 were collected
at 70\degc~due to the faster transition rate. Since the start and end phase of both temperatures
are the same, we use the first and last spectra collected at 70\degc~as the reference spectra 
used for the fitting residue term of the comprehensive acquisition
function, and use the one at 50\degc~for sampling test. 
Line plots of the normalized ground truth spectra at the beginning and end of the phase transition at both
temperatures are shown in Fig.~S1 of the Supplemental Document. 

To control the number of measurements per spectrum, we only used 10 randomly sampled initial points to
build the GP. We found this sparse set of initial points may leave out the relatively sharp pre-edge
peak, causing the lengthscale to be overestimated if it was found via maximum-likelihood fitting; therefore, we used a set lengthscale of 7 eV
based on our observation of the spectra. During an actual experiment, estimating the lengthscale
is also possible through observing the reference data while the ground truth of test data is unavailable. 
The maximum number of total measurements was set to 40. 

\begin{figure}
    \centering
    \includegraphics[width=0.8\linewidth]{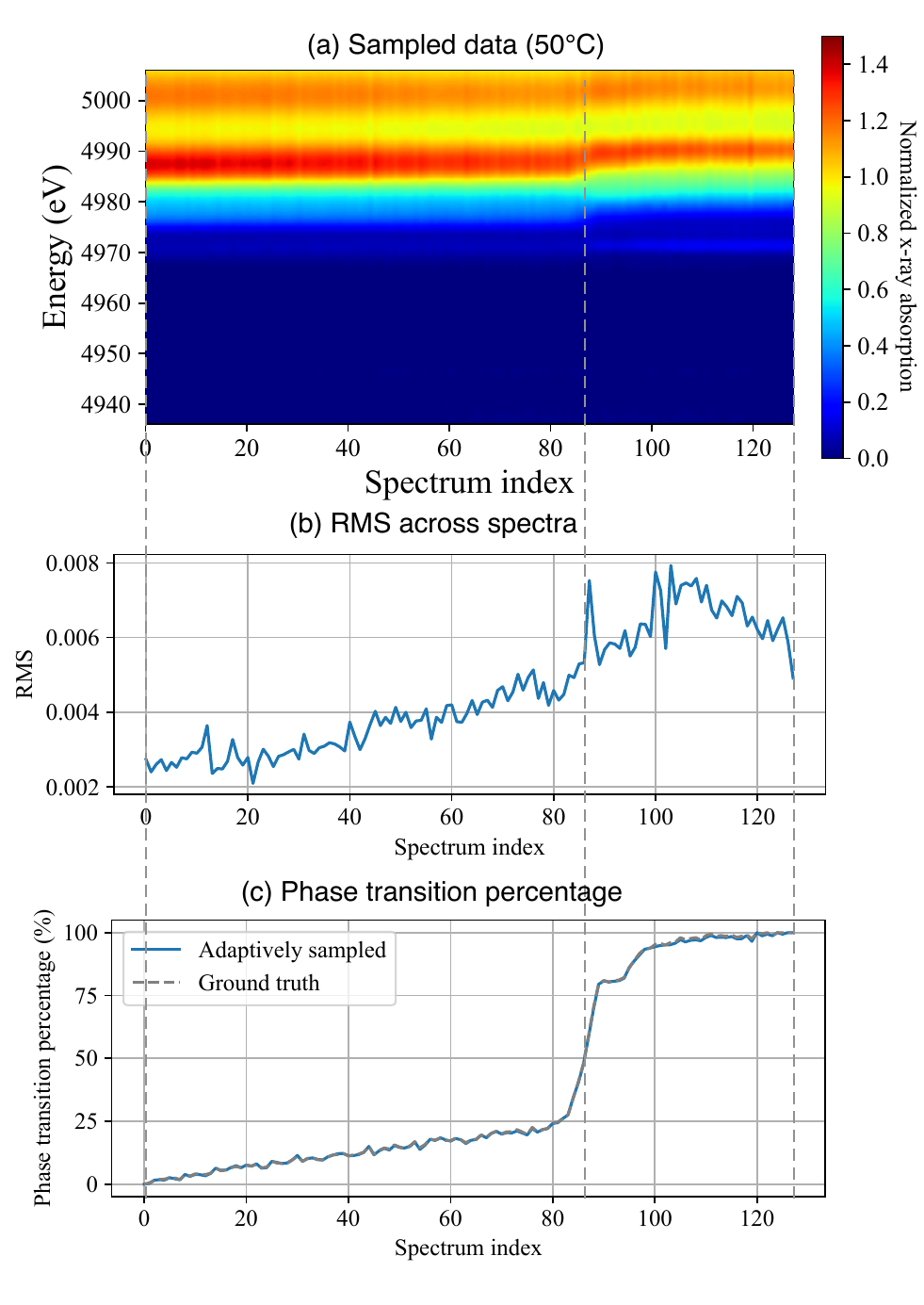}
    \caption{Results of applying our adaptive sampling method in a dynamic XANES experiment that tracks the
    phase transition progress of the LTO sample. (a) The stack of spectra at 50\degc~sampled using our method. Data plotted and normalized.
    (b) The RMS error of normalized spectra plotted against spectrum index. 
    (c) The phase transition percentages calculated through linear fitting with data collected. Percentages were calculated using normalized data. 
    (b) and (c) are aligned with (a) in the horizontal axis.
    The gray dashed lines marks the beginning, end of the experiment and the onset of the 
    phase transition.}
    \label{fig:lto_dynamic_experiment}
\end{figure}

Since the XANES spectrum evolves during a dynamic experiment, we run adaptive sampling
independently for every measured spectrum. With this approach, we sampled the 128 spectra in the test
set. 
The stack of sampled data is plotted in Fig.~\ref{fig:lto_dynamic_experiment}(a).
For presentation purposes, the data plotted are normalized. 
The RMS error between each reconstructed spectrum is calculated with the corresponding ground truth
after normalization, 
and is plotted in Fig.~\ref{fig:lto_dynamic_experiment}(b). The RMS error
remains around or below 0.003 before spectrum
85, after which it starts to increase, but stays below 0.008. This inflection point is marked by the
gray vertical dashed line across Fig.~\ref{fig:lto_dynamic_experiment}; the horizontal
axes of all the plots are aligned. In (b), the marked position shows an obvious change of spectral
structure, indicating the onset of phase transition. The phase transition results in a sharper
and higher pre-edge peak at 4972 eV which is the main source of the increased RMS error, but
overall this peak is still well captured in the reconstructed spectrum. The reconstructed spectra before
and after this transition (spectrum 80 and 90), before normalization,
are shown in Fig.~S2 in the Supplemental Document. 

Using the first and last spectra of the reference set as bases, we linearly fit each reconstructed
spectrum, and the obtained coefficients were used to estimate the percentage of phase transition
when that spectrum was measured. The trajectories of the transition percentage calculated
using the adaptively sampled spectra as well as the ground truth data are plotted
in Fig.~\ref{fig:lto_dynamic_experiment}(c). Despite the slightly elevated RMS error when the phase transition
begins, the percentages calculated using adaptively sampled spectra are very close to the ones
obtained with ground truth data: the maximum error in the percentage is only 1\%, while the linear combination fitting analysis typically has an uncertainty of a few percents. Considering that a maximum of 40 points are sampled for each
spectrum whereas
the ground truth data were collected with 141 points per spectrum,
our method reduced the number of measurements by 72\% without losing accuracy 
in the phase transition's kinetic information. 

\subsubsection{Sampling in a dynamic XANES experiment on the Pt/$\gamma$-Al$_2$O$_3$ sample}
 
In this section, we present a case study on sampling the XANES spectra of a 
0.35 wt\% Pt on $\gamma$-Al$_2$O$_3$ catalyst sample during an \insitu~reduction process. During the collection of the original data,
the sample was heated from from 26\degc~to 497\degc~in H$_2$, resulting in 
continuous reduction of the Pt oxidation state. Pt XANES measurements 
were recorded during this process. Details regarding the sample preparation 
and reduction conditions can be found in \cite{kelly_cjc_2019}. The strong peak at the 
absorption edge, also known as the white line, corresponds to the empty 5\emph{d} 
electrons of Pt$^{4+}$. As the Pt becomes reduced and these empty states become 
occupied, the XANES white-line intensity decreases \cite{chen_jpc_2021}. This sample 
is particularly interesting for this study due to the low Pt loading 
resulting in a small edge step and the large background absorption from 
the Al$_2$O$_3$ creating a large slope in the raw x-ray absorption spectra. 22 ground truth spectra were collected throughout the reduction process.
In the range between 11400 eV and 11700 eV, 238 equally spaced energy points were measured.

Compared to the YBCO sample, Pt's spectra at room temperature has a sharp 
and high white line, \emph{i.e.}, the absorption peak immediately
following the absorption edge. This poses a major challenge to adaptive sampling
as the accurate reconstruction of the white line requires at least one measurement
close enough to the peak's maximum, and any deviation would cause the white line's
height or position to be misrepresented in the reconstructed spectrum. The second-order
derivative and fitting residue terms in our comprehensive acquisition function
are particularly helpful in locating the peak's maximum, and as such, we specify
relatively high weight values for both terms ($\acqfweightgg = 2\times 10^{-3}$, 
$\acqfweightr = 100$). 
We take the first (26\degc) and last (497\degc) spectrum as the reference
data, and the remaining 20 spectra are used as the test set. 20 initial measurements are collected
randomly at the beginning of each run and the lengthscale parameter is fit on the initial data.

We start by examining the performance of our algorithm on a single spectrum. We choose
the spectrum measured at 42\degc~which features a high white line. Fig.~S3 in the Supplemental Document 
shows the sampling process at 20, 25, 30, and 50 total measured points. 
Our algorithm managed to precisely place a point at the maximum of the white line
after 5 additional measurements, and after that it starts to explore the uncertain regions. 
In addition to the high white line, Fig.~S3 also reveals a large
slope in the background due to the large absorption of the Al$_2$O$_3$ substrate,
but our algorithm still works robustly.
We also compare the RMS error convergence of our method (with and without
acquisition function reweighting) with BO using uncertainty-only acquisition function and
uniform sampling (Fig.~S4(a)).
The methods employing the comprehensive acquisition function clearly outperform
the latter two: an RMS error (on normalized spectra) of 0.005
is reached with about 40 total measurements, or
17\% of the number of ground truth points in the sampled range. 
On the other hand, 
uncertainty-only BO and uniform sampling still have high RMS errors above 0.05 
after 70 total measurements. The AUCs of the RMS convergence curves reveal the same trend, where the 5-run averaged AUCs of methods with
the comprehensive acquisition function is around 6 and 8 times lower than the other 2 methods.
The high errors of the latter mainly
originate from their failure to precisely capture the white line, as shown in
Fig.~S4(b) which compares the sampling process of the 4 methods. Similar to the LTO case, the difference between the AUCs with and without acquisition reweighting is not statistically significant, but a closer
examination around the white line's peak (inset in the last panel of Fig.~S4(b)) reveals more accurate sampling of such sharp features with reweighting. 

\begin{figure}
    \centering
    \includegraphics[width=1\linewidth]{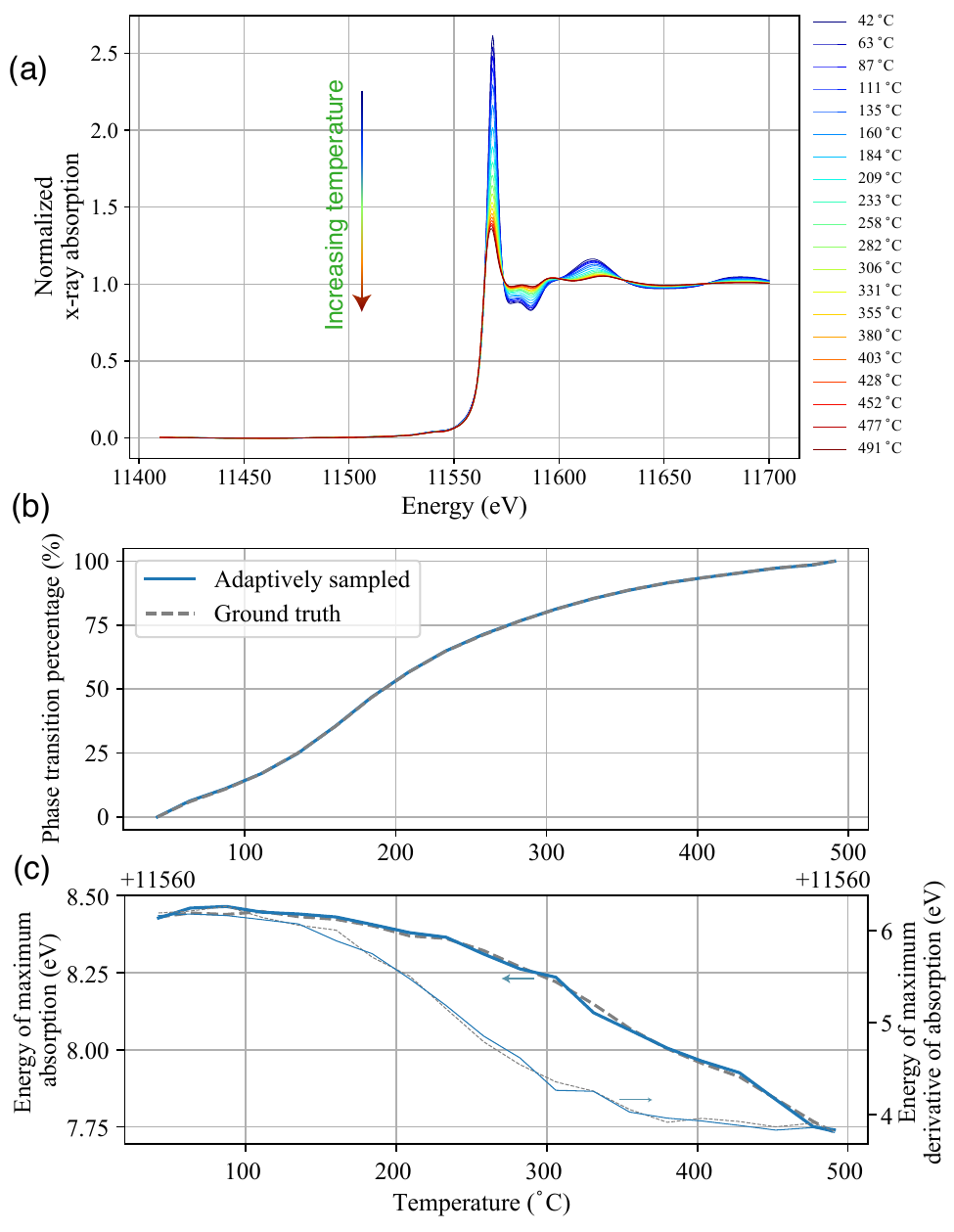}
    \caption{Dynamic experiment results of the Pt sample. 
    (a) All normalized spectra during 
    the reduction process, sampled using our method. (b) The transition percentages calculated using normalized data sampled with our method. 
    (c) The energies of the maxima of the white lines (bold lines), as well as those of the first derivative of the absorption edge (thin lines), plotted for all spectra over the reduction process. }
    \label{fig:pt_dynamic}
\end{figure}

We then test our method on data of the dynamic XANES experiment. Fig.~\ref{fig:pt_dynamic}(a)
plots the 20 spectra (excluding the reference spectra) sampled using our method, 
which are normalized for showing purpose. 
The percentages of state
evolution of the sample, calculated through the linear fitting to the normalized reconstructed
spectra using the two reference spectra as bases, are shown in Fig.~\ref{fig:pt_dynamic}(b).
Again, the percentage trajectories computed using adaptively sampled spectra
agree well with that calculated with the ground truth data, with the maximum error in the percentage being 0.3\%. 

The energy of the maximum of the white line, as well as the location of the absorption edge quantified by maximum point of the first derivative of the edge, are also important quantities that characterize the sample's chemical state. The Pt/$\gamma$-Al$_2$O$_3$ sample features a sharp white line and steep edge which make it an ideal case to assess the accuracy of our algorithm in determining both quantities. For each spectrum during the heating process, we fit a quadratic function to the local absorption values in a small neighborhood around the maximum of the spectrum to determine its location in energy. The same is done for the first derivative of the spectrum. Fig.~\ref{fig:pt_dynamic}(c) plots the evolution of the energy of the white line's maximum (thick lines) and that of the edge derivative's maximum (thin lines). The data for the adaptively sampled spectra (blue) and the ground truths (gray) are plotted in the same graph. Both quantities show good agreement, with a maximum error of less than 0.03 eV for the white line and less than 0.1 eV for the edge. The standard monochromator selects the incident x-ray energy with a FWHM of 1.3 eV at 10 KeV, so an error of less than 0.1 eV is smaller than our measurement resolution \cite{matsushita_handbook_1983}.

\subsection{Validation in a real experiment}

\begin{figure}
    \centering
    \includegraphics[width=1\linewidth]{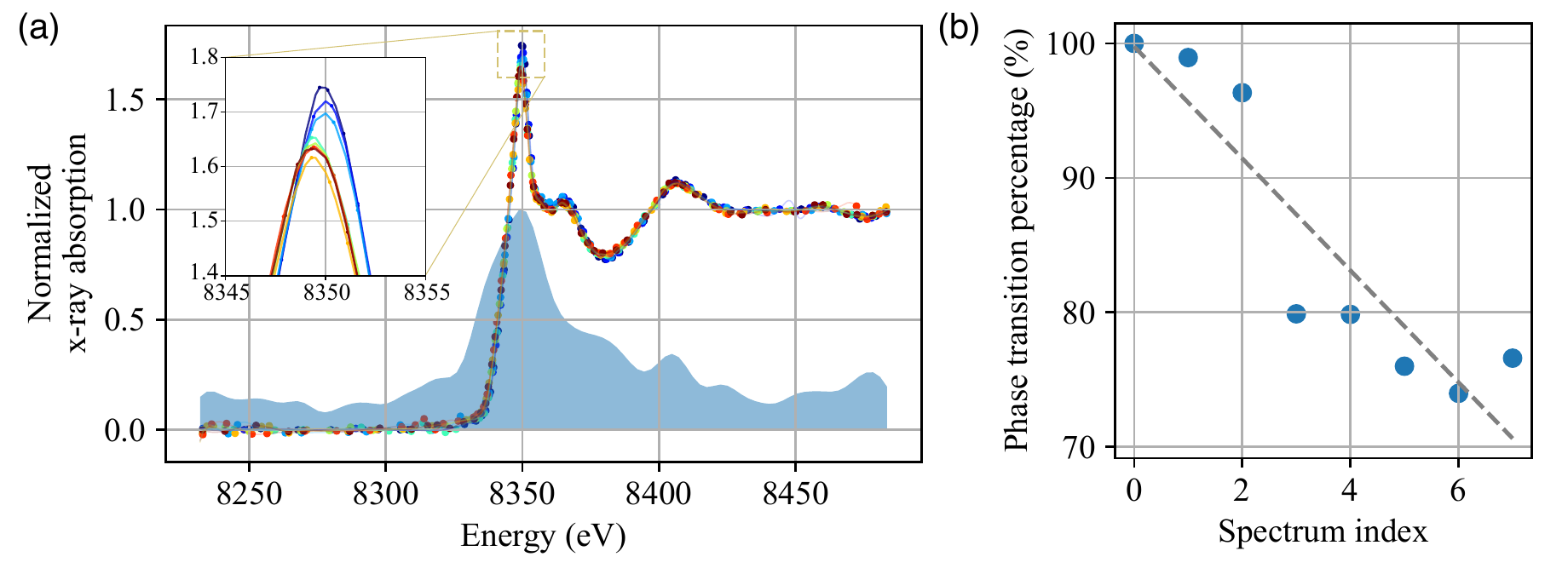}
    \caption{Dynamic experiment results of the NMC111 sample. 
    (a) The 8 normalized spectra during 
    the discharging process sampled using our method, plotted together and with the
    sampled points indicated by dots. From violet to red, the color of the lines
    and dots indicate the time sequence of the data. The densities of sampled points
    aggregated from all the 8 spectra are shown by the shaded region. 
    (b) The transition percentages calculated using normalized data sampled with our method as comparison of the Ni$^{2+}$ and Ni$^{3+}$ standards. The gray dashed line is the regression line.}
    \label{fig:nmc111_dynamic}
\end{figure}

We have deployed our algorithm at the 25-ID-C beamline at the Advanced
Photon Source and used it to guide a real-world experiment. 
The Bluesky infrastructure \cite{allan_synchrotron_rad_news_2019}
allows the adaptive sampling algorithm to receive the measured intensities
(and thereby the absorption coefficients), and send suggestions
of the next energy to measure. With this control setup, we performed a smartly controlled
in-situ XANES experiment that measures the absorptions of a
lithium nickel manganese cobalt oxide (\ce{LiNi_{1/3}Mn_{1/3}Co_{1/3}O2}) battery electrode during its discharging
process. We first collected 2 reference spectra on a pristine sample and
a fully charged (4.5 V) sample
in the range between -100 eV to +100 eV relative to the Ni K-edge of 8333 eV. We then sequentially measured 8 spectra under the guidance
of our algorithm as the in-situ cell was discharged from 4.5 V.
To make the case more challenging for our algorithm, the adaptive spectra
were collected in a larger range of -100 eV to +150 eV relative to the Ni
edge, \emph{i.e.}, no reference data is available for the last 50 eV.
However, our algorithm still performed robustly. 
The normalized traditional spectra are plotted
together in Fig.~\ref{fig:nmc111_dynamic}(a), with the sampled points indicated
by markers. The first spectrum collected in the process is plotted in violet,
and the last is in red. We also calculated the sampling density of all energies
sampled among the 8 spectra using kernel density estimation, and plot it as the
shaded area at the bottom of the figure. The sampling density is highest at the absorption
edge and the white line, which are the most important features to track the change
of oxidation state of an element during a dynamic process. Fast-varying features,
like peaks and valleys in the post-edge region, are also more densely sampled.
For example, the spectrum peak at 8405 eV is well aligned with a peak in the
sampling density. smoothly varying regions away from the edge are sampled less
to reduce the number of measurements. 

In terms of experiment time, traditional densely collecting a reference 
spectrum of 401 points took 620 seconds on average. Considering that the
energy range of the reference spectrum is 4/5 of the adaptively sampled range,
we scale it to estimate the time for densely sampling the 250 eV's range, which gives
775 seconds. The adaptive sampling
of 60 points, including the 20 initial measurements and the subsequent 40 adaptive
measurements, took an average of 180 seconds, or 23.2\% of dense sampling. This ratio
is higher than that of the numbers of points (12.0\%), but is still in a reasonable range considering the overhead of computation and the longer average travel distance of the motors to position the monochromator from the current energy to the next.

Using the reference spectra as bases, we estimate the percentage of Ni's oxidation as it reduces from around +4 to +2,
as shown in
Fig.~\ref{fig:nmc111_dynamic}(b). The plot reveals the decreasing trend in
the oxidation state, expected for a discharging process; it also reveals that
the reduction of Ni was not complete, which plateaus at about 75\% or approximately
+3.5. This observation provides insights about the electrode's cycling
mechanisms for battery researchers. 

We summarize the 4 test cases by listing the numbers of points measured adaptively 
(for the 3 simulated
cases, they are the numbers of points to reach an RMS error of 0.005 compared
to the ground truths) and
densely sampled in Table \ref{tab:n_points}. The ratios of the numbers of points sampled adaptively
and densely sampled are also reported.

\begin{table}[]
    \centering
    \begin{tabular}{cccccc}
    \hline
        \textbf{Sample} & \textbf{Type} & \textbf{Range (eV)} & \makecell{\textbf{\# points}\\ \textbf{(adaptive)}} & \makecell{\textbf{\# points} \\ \textbf{(traditional)}} & \makecell{\textbf{Adaptive} \\ \textbf{sampling} \\ \textbf{ratio} (\%)} \\
        \hline
        YBCO & Simulated & 160 & 34 & 218 & 15.6 \\
        LTO & Simulated & 70 & 40 & 141 & 28.4 \\
        Pt & Simulated & 300 & 40 & 238 & 16.8 \\
        NMC111 & Real & 250 & 60 & 501 (scaled) & 12.0 \\
        \hline
    \end{tabular}
    \caption{The number of measured points of adaptive and dense sampling, and their
    ratios, for cases shown in this paper. Since the reference spectra of the NMC111
    case were collected in a smaller range (200 eV) than the adaptively sampled
    spectra (250 eV), we scale the actual number of points in reference spectra (401)
    to match the range of adaptively sampled spectra.}
    \label{tab:n_points}
\end{table}

\section{Discussion}

\subsection{Comparison to other approaches}

Our adaptive sampling algorithm is based on GP-BO and 
leverages domain-knowledge specific to the structure of XANES
spectra to guide the experiment such that important features are sampled more frequently. 
This knowledge is injected by designing an acquisition functions that returns higher values
where (1) fast variations, peaks, or valleys are present, (2) the reconstructed spectrum is not well
explained by the linear combination of reference spectra, and (3) the location is away from the smoothly varying
region below the absorption edge. Artificially designing the acquisition function for XANES 
is straightforward because the measured space is only one dimensional, and assumption that a XANES spectrum
sequentially contains a smooth pre-edge region, pre-edge peaks, an absorption edge, and post-edge features
generally holds true. Additionally, the fact that an XANES spectrum is an average of the element-of-interest oxidation states, it can be
represented by the linear combination of a few reference spectra providing an effective analytical
approach to assess the accuracy of the reconstructed spectrum. Therefore, our method works well
without the need of extensive training data as in the case of data-driven approaches like 
reinforcement learning \cite{betterton_icra_2020} or SLADS-Net \cite{zhang_electronicimaging_2018},
and is less susceptible to the generalization gap that data-driven methods often encounter
when working with unseen test data. 
Our method also provides an explicit uncertainty quantification and the ability to guide the 
acquisition using it compared to the two data-driven approaches listed
above. However, we would like to note that uncertainty quantification is also possible with
data-driven approaches. One such example is deep kernel learning (DKL) \cite{wilson_arxiv_2015}.
In DKL, the joint distribution of past and new observations is also modeled as a GP that
uses a common (and possibly stationary) kernel function, but 
a neural network $\deepkernel$ is trained project the input features to a different space
before they are processed by the kernel. In other words, the evaluation of the kernel is
modified to $\kernelfunc\left[ \deepkernel(\inputvar_1), \deepkernel(\inputvar_2) \right]$.
The non-linear projection of $\deepkernel$ allows the kernel to adapt to different feature
lengthscales in the data: points in the fast varying parts of the data can be projected
to have larger spacing which is equivalent to reducing the lengthscale specifically in that region. 
However, $\deepkernel$ is not necessarily shift-equivalent; when applied to XANES, that means
if the absorption edge's location in a test spectrum is different from those in the training data,
$\deepkernel$ does not recognize this difference by itself and would still project input features
based on their absolute positions from what it learns during training. This issue can be partly
solved by shifting the training and test data in such a way that their absorption edges are all at
a standardized position. This is easy to do for training data, but for test data the only way to
find the correct standardizing shift is to detect the edge's location using an sparse set of
initial measurements, which may be inaccurate. Additionally, since the structures and locations
of fine features vary drastically across different samples, it is hard for $\deepkernel$ to learn
a projection with sub-feature level precision. On the other hand, our method adaptively detects
peaks, valleys and high-gradient areas using derivatives and under-described areas using
fitting residues and escalates the acquisition function precisely in those areas. It might be
possible to adapt $\deepkernel$ in DKL to also take these computed features into account instead of
just the positions in the input space, but this is beyond the scope of this paper. 

\subsection{Computation cost}

A general concern about GP-BO is the $O(n^3)$ scaling with the number of measured points 
in inverting the covariance matrix for posterior calculation. 
However, the one dimensional measurement space of XANES eases this concern as the size of the covariance
matrix is well under control: the size of the matrix equals the number
of measurements, and in our test cases shown above, 
measuring an energy range of 100 - 300 eV typically takes just 40 - 60
measurements. 
On a CPU, the time taken by our algorithm to make a prediction is
below 0.1 second and the speed did not change significantly during the acquisition. 

\subsection{Limitations and future works}

A limitation of our method, which is also common to GP-BO algorithms using common kernel functions,
is that when the actual feature lengthscale in the data is extremely non-uniform, the
posterior mean in sparsely sampled smoothly varying regions may be inaccurate. 
This is explained by Eq.~\ref{eq:posterior_mean}
which shows that the posterior mean's residue from the prior mean is an average of
that residue of past measurements weighted by $\traintestcovar^T(\covar + \noisecovar)^{-1}$.
If the distance of a test point $\newinputvar$ to nearby past measurements is well above the kernel's
lengthscale, its covariance with these points is underestimated and points lying farther away,
including those with a very different observed values, become more highly weighted and
the posterior mean biases towards the weighted global average. As a result, it is likely
for the posterior mean in undersampled pre-edge smoothly varying region to oscillate. We note, however,
that this typically occurs when the measured energy range is very large and contains a
substantial portion of smoothly varying regions (more than 100 eV). One way to address this is to design
a projection function to map the input features to another space that better describes
the spatially varying lengthscale, in a way similar to DKL but with the projection function
designed analytically. However, a more straightforward workaround that we currently use in our algorithm 
is to reconstruct the spectrum using spline interpolation instead of the posterior
mean of GP, which avoids the oscillation. Additionally, the range of feature-rich region in an
XANES spectrum is typically known for a specific type of sample, so it is possible to
confine the sampling in that region without including an exceedingly large portion of
plateau. Even if the feature-rich range is not known,
we may automatically determine that during an experiment by first running
a series of coarse measurements across the energy range, 
and then estimating the absorption edge's location
using the energy-domain gradient of the measured values. Following that, the range for 
sampling can be chosen to be a few tens or hundreds of eVs 
below and above the edge energy. Nevertheless, although spline interpolation guarantees smooth reconstruction, a good lengthscale parameter is still needed to make the uncertainty component in the acquisition function work properly. We plan to explore the feature projection approach mentioned above to address this problem. 

Beyond overcoming the current limitations, our future prospect also includes exploring the broader applications of features in the algorithm. Notably, the linear fitting of the reconstructed spectrum with reference spectra, used in the fitting residue component of the comprehensive acquisition function, is also a technique commonly used in the analysis of XANES spectra, and the fitted coefficients reveals the progress of a dynamic process. At the end of measuring each spectrum, this progress information can be computed and reported to the user (or a higher-level experiment control algorithm), so that the whole experiment can be stopped immediately when the process finishes, avoiding taking redundant spectra. Moreover, the fitting also indicates how well the current spectrum is represented by the references. If the fitting residue of certain regions remains high even though these regions have been measured sufficiently, it most likely indicates that the reference spectra are improper for the sample being measured, and the user or the experiment control algorithm can be prompted to look for better reference spectra from the database. To summarize, beyond automating the collection of individual spectra, our algorithm could also become the core of a larger autonomous system that streamlines data collection and analysis. 

To conclude, we have developed an adaptive sampling 
algorithm for XANES spectroscopy based on Bayesian 
optimization. Domain knowledge about XANES is incorporated 
into the algorithm through a designed acquisition function 
that makes the algorithm aware of regions of higher 
interest in the energy domain in real time. With this, the
algorithm can direct experimental instruments to sample 
more densely at the absorption edge and other regions with 
fast varying features than areas that are more uniform. 
Through simulated studies and a real-world demonstration at 
the Advanced Photon Source, we show that our algorithm 
yields spectra as accurate as those sampled using 
traditional methods while only taking 10 -- 30\% of the 
measurements. The algorithm shortens the acquisition time 
of XANES, allowing for better time resolution in dynamic 
experiments, and avoids the over/undersampling problems in 
traditional XANES acquisition with pre-defined sampling 
grid. It signifies a step towards future x-ray experiment 
endstations that are more efficient, more autonomous, and 
more intelligent.



\section{Methods}
\label{sec:methods}

\subsection{Gaussian process and Bayesian optimization}

Standard Gaussian process (GP) and Bayesian optimization (BO) has been documented in a number of 
literature \cite{berger_statsbook_1985, garnett_bayesoptbook_2023}. We reiterate the fundamental
theories here in the context of the adaptive sampling problem under concern. In a spectroscopy
experiment like XANES, one measures a function $\spectrofunc(\inputvar)$ 
in the one-dimensional energy domain. For a series of $\numpastmeas$ past measurements collectively
denoted by $\pastmeas$ at locations $\pastlocs$, the probability density function of $\pastmeas$
is
\begin{eqnarray}
    p(\pastmeas) &=& \frac{1}{\sqrt{(2\pi)^\numpastmeas |\covar|}}\exp\left[ -\frac{1}{2}(\pastmeas - \mean)^T\covar^{-1}(\pastmeas - \mean) \right] \\ 
    &=& \mbox{GP}(\pastmeas; \mean, \covar).
    \label{eq:past_meas_gp}
\end{eqnarray}

Eq.~\ref{eq:past_meas_gp} is a GP describing the distribution of $\pastmeas$,
where $\mean$ is the expectation that often takes a simple form such as a constant. $\covar$ is the
covariance matrix with each element $K_{ij}$ representing the covariance between the observations
at $\inputvar_i$ and $\inputvar_j$. This covariance is estimated by the kernel function 
$k = \kernelfunc(\inputvar_i, \inputvar_j)$, where $\kernelhp$ is the set of hyperparameters of the
kernel. In this work, we use a \matern~kernel which takes the
following form:
\begin{equation}
    \kernelfunc(\inputvar_i, \inputvar_j) = \frac{1}{\Gamma(\nu)2^{\nu - 1}}\left( \frac{\sqrt{2\nu}}{l}|\inputvar_i - \inputvar_j| \right)^\nu K_\nu\left( \frac{\sqrt{2\nu}}{l}|\inputvar_i - \inputvar_j| \right)
    \label{eq:matern_kernel}
\end{equation}
where $K_\nu(\cdot)$ is a modified Bessel function and $\Gamma(\cdot)$ is the gamma function.
We use $\nu = 2.5$ because it yields overall the best performance on the data tested.
$l$ is the lengthscale and is the only tunable hyperparameter of the \matern~kernel, \emph{i.e.},
$\kernelhp = \{ l \}$. 

We further assume that the measurement is noisy so the actual past observations 
$\pastobs$ follows $p(\pastobs) = \mathcal{N}(\pastmeas, \noisecovar)$, 
where $\mathcal{N}$ is a Gaussian distribution. This allows us to formulate the
log-likelihood function as 
\begin{equation}
\begin{aligned}
    \log p(\pastobs | \pastlocs, \kernelhp) &= \log \int p(\pastobs | \pastmeas, \pastlocs, \kernelhp)p(\pastmeas | \pastlocs, \kernelhp)d\pastmeas \\
    &= -\frac{1}{2}\bigg[ (\pastobs - \mean)^T (\covar(\kernelhp) + \noisecovar)^{-1}(\pastobs - \mean) + \log|\covar(\kernelhp) + \noisecovar| \\
    &\quad + \numpastmeas\log(2\pi) \bigg].
    \label{eq:likelihood}
\end{aligned}
\end{equation}

By maximizing Eq.~\ref{eq:likelihood} with regards to $\kernelhp$, one can fit the kernel
function's hyperparameters when constructing a GP model 
using an initial set of observations and their locations. However, if one has good knowledge
about the sample being measured, it is also possible to set strong priors or fixed values
for the hyperparameters. This is applicable to the lengthscale parameter of the 
\matern~kernel, where we have found setting the lengthscale with a good estimate sometimes
leads to better performance than fitting it on initially measured data when the initial
points are scarce.

When a new measurement is made at point $\newinputvar$, the joint distribution of the measured value
$\newmeas$ and previous measurements $\pastmeas$ is
\begin{equation}
    p(\newmeas, \pastmeas) = GP\left(
        \begin{bmatrix}
            \newmeas \\
            \pastmeas
        \end{bmatrix};
        \begin{bmatrix}
            \mu^* \\
            \mean
        \end{bmatrix},
        \begin{bmatrix}
            K^* & \traintestcovar \\
            \traintestcovar^T & \covar
        \end{bmatrix}
    \right)
\end{equation}
where $K^* = \kernelfunc(\newinputvar, \newinputvar)$, $\traintestcovar = \kernelfunc(\newinputvar, \pastlocs)$.

Taking observation noise into account, 
the posterior predictive distribution of $\newmeas$ at $\newinputvar$ is
\begin{equation}
    p(\newmeas | \pastobs) \propto \mathcal{N}\left[ \posteriormean(\newinputvar), \posteriorvariance(\newinputvar) \right]
    \label{eq:posterior_predictive}
\end{equation}
with
\begin{equation}
    \posteriormean(\newinputvar) = \mean + \traintestcovar^T(\covar + \noisecovar)^{-1}(\pastobs - \mean)
    \label{eq:posterior_mean}
\end{equation}
and
\begin{equation}
    \posteriorvariance(\newinputvar) = \kernelfunc(\newinputvar, \newinputvar) + \traintestcovar^T(\covar + \noisecovar)^{-1}\traintestcovar.
    \label{eq:posterior_variance}
\end{equation}

Eq.~\ref{eq:posterior_predictive} -- \ref{eq:posterior_variance} provides a value estimation 
(through the posterior mean $\posteriormean(\newinputvar)$) and uncertainty quantification
(through the posterior variance $\posteriorvariance(\newinputvar)$) for $\newmeas$. Based on these
quantities, one may create an acquisition function $\acqf(\inputvar)$ that evaluates the 
``worth'' of taking a measurement at any point $\inputvar$. The formulation of $\acqf(\inputvar)$
can be designed to reflect the objective of the experiment. For example, if the objective is to
reduce the overall uncertainty of the data, then $\acqf(\inputvar)$ could simply be the
posterior standard deviation, \emph{i.e.}, $\acqf(\inputvar) = \posteriorstd(x)$. 

With the acquisition function defined, 
one can guide subsequent data acquisition through a Bayesian optimization workflow:
one finds the maximum point of $\acqf(\inputvar)$, \emph{i.e.}, 
$\newinputvar = \argmax_\inputvar \acqf(\inputvar)$, and take the next measurement at $\newinputvar$.
$\newinputvar$ and the measured value $\newmeas$ are used to update $\pastobs$, $\mean$, $\traintestcovar$, and $\covar$, yielding the new posterior predictive distribution parameters (Eq.~\ref{eq:posterior_mean}
and \ref{eq:posterior_variance}). This process is repeated until a stopping criterion is met. 

\subsection{Acquisition function design for XANES adaptive sampling}

From Eq.~\ref{eq:posterior_variance}, one can find that the posterior variance is only a function of the
\emph{locations} of past measurements, but not their values; furthermore, a stationary kernel
like \matern~is only dependent on the distance rather than absolute positions of the two input points, 
and so is the posterior variance. It is then obvious that simply using the posterior variance
in the acquisition function $\acqf(\inputvar)$ has the drawback of not being aware of the absolute positions
and values of the measured or GP-estimated spectrum (\emph{i.e.}, its posterior mean). To allow
the adaptive sampling algorithm to make better decisions in the points to measure, we inject our domain
knowledge about XANES spectra by designing the acquisition function in a more robust way. 

\subsubsection{Gradient component}

In an XANES spectrum, the most important features are the x-ray energy of the absorption edge, the pre-edge features and undulations
around the absorption edge. The edge itself and the inclining and declining parts of wavy features
can be identified by the large magnitude of their gradient with regards to energy. The peaks
and valleys of near-edge undulations on the other hand are highlighted by their relatively large
second-order gradient. Thus, we take both the magnitudes of the 
first- and second-order gradient of the posterior
mean $\posteriormean(\inputvar)$ in formulating the gradient component of the acquisition function:
\begin{equation}
    \acqfg(\inputvar) = \acqfweightg\left\|\frac{d\posteriormean(\inputvar)}{d\inputvar}\right\|
        + \acqfweightgg\left\|\frac{d^2\posteriormean(\inputvar)}{d\inputvar^2}\right\|
    \label{eq:acqf_grad}
\end{equation}
where $\acqfweightg$ and $\acqfweightgg$ are weight coefficients. The values of these coefficients
can be scheduled to gradually reduce throughout the experiment, so that the algorithm focuses on
regions with high first or second-order derivatives at first but slowly shifts towards exploration
on uncertainty regions. We use an exponential decaying schedule for these coefficients. Each
time the GP model is updated with new measurement, the weight coefficients are updated as
\begin{equation}
    \phi_i \leftarrow \beta\phi_i \qquad (i \in \{ g, g' \})
    \label{eq:exponential_decay_schedule}
\end{equation}
$\beta$ is set to be 0.999.

We also allow the initial values of $\acqfweightg$ and $\acqfweightgg$ to be estimated automatically as
\begin{equation}
    \phi_i = \alpha_i\frac{\max{\posteriorstd(\inputvar)}}{\max a_{g,i}(\inputvar, \phi_i = 1)} \qquad (i \in \{ g, g' \})
    \label{eq:acqf_weight_coeff_estimate}
\end{equation}
where $\alpha_i$ is the importance relative to the posterior uncertainty 
for term $i$ (first-order or second-order), 
and $a_{g,i}$ refers to the corresponding term in $\acqfg$. We set both $\alpha_i$'s to 0.5. 
Eq.~\ref{eq:acqf_weight_coeff_estimate} scales the acquisition function component's values
so that their maxima are comparable with the posterior standard deviation.

Lastly, considering that the XANES spectra measured by a realistic instrument sometimes contain
a background slope which may bias the first-order gradient, we detect the range of the pre-edge
region using the method described in Section \ref{sec:reweighting} and estimate the gradient in
that region. This allows us to subtract the bias from the first-order gradient.

\subsubsection{Fitting residue component}

In dynamic XANES where multiple spectra of the sample are measured at different time points during
a evolving process (\emph{e.g.}, phase transition), the XANES spectra
of the equilibrium-state phases before and after the transition are often 
available as reference spectra.
A spectrum measured during phase transition can be accurately described by the linear 
combination of the reference spectra \cite{kelly_2015}. Therefore, during a guided
experiment, if one fits the posterior mean $\posteriormean(\inputvar)$ (\emph{i.e.}, the densely reconstructed spectrum)
with the reference spectra, a high fitting residue would indicate that values estimated for 
that local region might have deviated from the true values, and more points should be measured
there. We thereby formulate the fitting residue component of the acquisition function as
\begin{equation}
    \acqfr(\inputvar) = \acqfweightr\left\| \fittedspectrum(\inputvar) - \posteriormean(\inputvar) \right\|^2
    \label{eq:acqf_residue}
\end{equation}
where $\fittedspectrum(\inputvar)$ is spectrum linearly fitted to $\posteriormean(\inputvar)$
using the reference spectra as bases. Like in the case of the gradient component, the weight
coefficient $\acqfweightr$ can also be scheduled to decay exponentially. The initial values
of $\acqfweightr$ may also be automatically determined using a method similar to the 
gradient component.

\subsubsection{Acquisition function reweighting}
\label{sec:reweighting}

Both the gradient and fitting residue components are calculated based on $\posteriormean(\inputvar)$,
which may deviate from the actual data. For example, a low-gradient region rendering a small $\acqfg$
might in fact contain a small feature not detected yet. Therefore, we still need to keep the
posterior standard deviation $\posteriorstd$ in the acquisition function to allow the algorithm
to explore uncertain regions even if $\acqfg$ and $\acqfr$ are low there. This illicits another
issue that the algorithm might allocate more-than-enough points in the smoothly varying pre-edge region
driven by its high uncertainty. Hence, we create another modifier for $\acqf(\inputvar)$ to reweight
its value across the measured space. 

Given the $\posteriormean(\inputvar)$ computed after the GP model is updated with a few measurements,
we find the location $\edgeloc$ and width $u$ of the absorption edge in the 
posterior estimate by taking its
finite-difference derivative with regards to energy, and identifying the largest peak of the
derivative. This allows us to estimate the upper bound of the smoothly varying pre-edge region as 
$\inputvar_{pe} = \edgeloc - c_e u$, where $c_e$ is an adjustable but widely applicable constant that is set to be
1.6 in all tests performed. A shifted, scaled, and elevated sigmoid function is then constructed as
\begin{equation}
    w_s(\inputvar) = \left( \frac{1}{1 + \exp\left[ -\frac{3200}{u}(x - \inputvar_{pe}) \right]}
     \right)(1 - w_{s,\text{floor}}) + \weightfloor.
    \label{eq:sigmoid_weight}
\end{equation}
The units of $\inputvar$, $\inputvar_{pe}$ and $u$ are eV. $\weightfloor$ is the floor or
infimum value of the sigmoid, which is set to a small but non-zero number. 
When $\weightfloor$ is multiplied to
the $\acqf(\inputvar)$, it scales down the acquisition function's value in the pre-edge region by
$\weightfloor$, but without completely zeroing it out. This allows the algorithm to sample more frequently
in the edge and post-edge region while maintaining the possibility of uncertainty-driven
exploration in the pre-edge region when the edge and post-edge acquisition function become low
enough after sufficient amount of points are sampled. The factor
of 3200 is determined empirically. 

\begin{figure}
    \centering
    \includegraphics[width=0.7\linewidth]{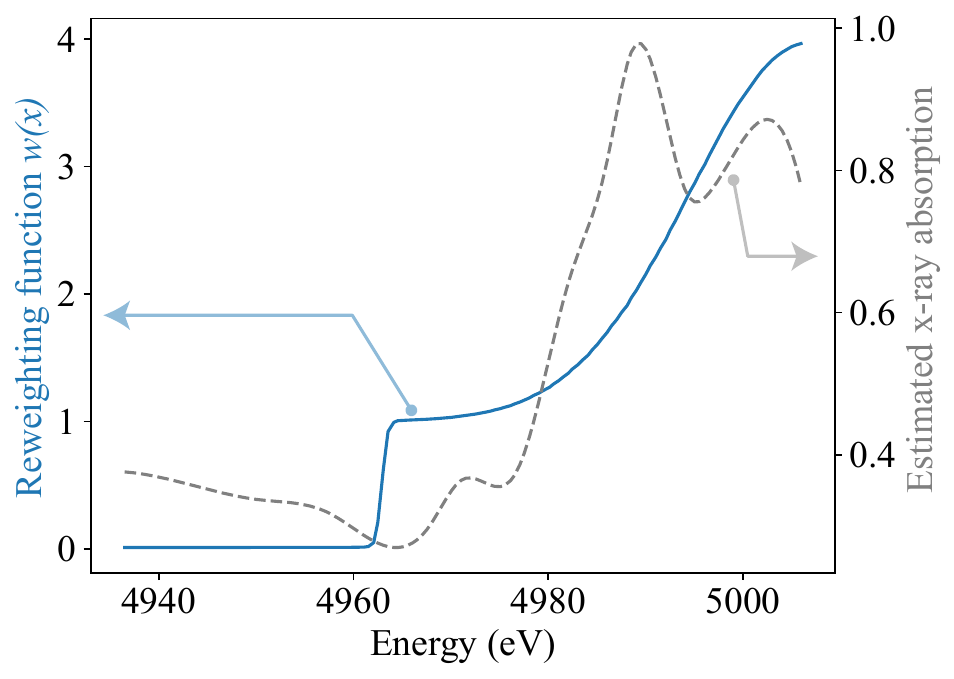}
    \caption{An example acquisition reweighting function $w(\inputvar)$ (blue, left axis)
    plotted along with the x-ray absorption estimated through posterior mean (gray, right axis).}
    \label{fig:acquisition_reweighting}
\end{figure}

Optionally, one may also boost up the acquisition function in the region closely after 
the absorption edge, which contains most of the informative features.
This is done by adding a Gaussian function of
\begin{equation}
    w_g(\inputvar) = g\exp\left( -\frac{[\inputvar - (\edgeloc + u)]^2}{0.5u^2} \right)
    \label{eq:gaussians_weight}
\end{equation}
where $g$ is a gain value. Together, the acquisition reweighting function is
\begin{equation}
    w(\inputvar) = w_s(\inputvar) + w_g(\inputvar).
    \label{eq:acquisition_weight}
\end{equation}
An example reweighting function is shown together with the posterior mean used to generate it in Fig.~\ref{fig:acquisition_reweighting}.

The acquisition function is modified by this reweighting function as the blending between
the original acquisition function and the weighted version of it. If the unmodified acquisition
function is $\acqf_0(\inputvar)$, then the modified function is given as
\begin{equation}
    \acqf(\inputvar) = \gamma w(\inputvar) \acqf_0(\inputvar) + (1 - \gamma)\acqf_0(\inputvar)
    \label{eq:blended_acqf}
\end{equation}
where $\gamma$ is the blending coefficient which can also be scheduled to decay from the initial
value of 1.0 so that the effect of reweighting gradually fades out, letting uncertainty
play a more significant role at later stages. We use an exponential decay schedule
similar to the weight coefficients of the acquisition function components and the decay
coefficient is set to 0.95 unless otherwise specified. 

\subsubsection{Comprehensive acquisition function}

With the features introduced above, we create a comprehensive acquisition function that
is value-aware and incorporates our domain knowledge about XANES spectra. The acquisition
function is formulated as the following:
\begin{equation}
    \acqf(\inputvar) = w(\inputvar)\posteriorstd(\inputvar)\max(\acqfg + \acqfr, \tau)
    \label{eq:comprehensive_acqf}
\end{equation}
where $\tau$ is a lower bound that is set to be 0.03 throughout this work unless otherwise
mentioned. The posterior variance $\posteriorstd(\inputvar)$ and domain knowledge components
($\acqfg(\inputvar)$ and $\acqfr(\inputvar)$) are multiplied instead of added. This is based
on the fact that when a point is measured, the $\posteriorstd(\inputvar)$ in its close
proximity drops towards 0. We want this behavior to control the entire acquisition function,
so that the algorithm will not make repeated measurement in a neighborhood where enough
points have already been measured. Also, the sum of domain knowledge components
$\acqfg + \acqfr$ are constrained with a lower bound 
$\tau$, so that a seemingly smoothly varying or low-residue region
does not totally zero out the acquisition function, and the algorithm preserves the possibility
to take measurements there driven by uncertainty. These two designs prevents the algorithm
from either over-sampling in a region or missing regions containing undetected features. 

\subsubsection{Early stopping}
\label{sec:early_stopping}

The acquisition can be stopped when the maximum posterior standard deviation 
weighted by the sigmoid term of the reweighting function drops below a threshold, \emph{i.e.},
\begin{equation}
    \max\left[ \posteriorstd(\inputvar)w_s(\inputvar) \right] < t.
    \label{eq:early_stop}
\end{equation}

\subsubsection{Maximization of the designed acquisition function}

Many BO algorithms find the maxima of the acquisition function through numerical optimization. However, since the input space of XANES is one dimensional, it is in fact more convenient and effective
to numerically find the maximum of the acquisition function by evaluating it on a fine grid of
points, and then simply taking the maximum. 

\subsection{Spectrum reconstruction}

The data collected with adaptive sampling are typically sparse and non-uniform. Eventually, one will
need to recover a spectrum with its values interpolated on a denser grid in the energy domain.
This reconstruction is naturally available through a GP model, 
where one may evaluate the posterior
mean $\posteriormean(\inputvar)$ at a given set of energies $\inputvar$, and use these values
as the reconstructed spectrum. 

However, reconstructing the spectrum using posterior mean may give suboptimal results under 
some special circumstances. With the use of the comprehensive acquisition function, 
sampled points in the pre-edge regions 
of the spectra might be particularly sparse when the experiment stops. The spacings between
measured points in those regions are likely significantly greater than the kernel function's
lengthscale. When this is the case, the correlation of such points to the measured points nearby
is underestimated, and the posterior mean $\posteriormean(\inputvar)$ at these points biases
towards a globally averaged value of all past observations according to Eq.~\ref{eq:posterior_mean}.
As a consequence, the spectrum reconstructed using $\posteriormean(\inputvar)$ tends
to oscillate at these regions while it is supposed to be smooth. One can work around this by,
for example, non-linearly mapping the measured domain to a different manifold such that the
spacings between points in pre-edge regions are smaller after mapping \cite{wilson_arxiv_2015}.
However, here we take a simpler approach, where instead of reconstructing the spectrum using
posterior mean, we use the cubic spline interpolation of measured points. This results in a
smooth connection among the points that has little deviation from the posterior mean at regions
adequately described by the kernel lengthscale, while avoiding the spurious oscillation in
smoothly varying regions.

\subsection{Experimental data collection}

Our algorithm is tested with data collected on 4 samples, namely a superconductor material
yttrium barium copper oxide (YBCO), a battery material lithium titanium oxide (LTO), 
a catalytic material
with Pt particles deposited on $\gamma$-Al$_2$O$_3$ substrate, and finally
a battery material lithium nickel manganese cobalt oxide (NMC111). 
The demonstrations of our algorithm on the first 3 are simulated, meaning 
that ground truth spectra have been collected using a dense traditional sampling grid beforehand,
and the algorithm runs with data interpolated from the truths. The test on the
NMC111 sample was done in the real world, with the algorithm sending and receiving
data to/from the monochromator and photon detector. This section
documents the experimental collection of the data. 

\subsubsection{Data collection of the LTO and YBCO samples}

The Li$_4$Ti$_7$O$_{12}$ (LTO) samples were commercially purchased from EnerDel. 
The preparation of orientation-dependent YBa$_2$Cu$_3$O$_7$ (YBCO) samples involves 
synthesizing YBCO powder and orienting the powders with a high magnetic field, 
the details of which can be found in \cite{heald_physrevb_1988}. 

The XANES spectra of both YBCO and LTO were measured 
at the 20-BM beamline \cite{heald_jsr_1999} of the Advanced Photon Source (APS). 
In both experiments, a harmonic rejection mirror and a 10-15\% detuning of x-ray 
intensity using the 2nd crystal of the Si (111) monochromator were applied to 
avoid high-energy x-ray harmonics. These measurements were performed in transmission mode.
Two LTO samples were respectively measured at 2 different temperatures, 50\degc{} and 70\degc{}.
At each temperature, multiple XANES spectra were collected throughout the sample's
phase transition process. 
The Linkam THMS600 heating stage was used for temperature-dependent measurements. 

\subsubsection{Data collection of the Pt/$\gamma$-Al$_2$O$_3$ sample}

For the data collection of the Pt/$\gamma$-Al$_2$O$_3$ material, a sample was prepared following
the procedures in \cite{kelly_cjc_2019}. XANES spectra were continuously measured on
the sample while it undergoes reduction in hydrogen atmosphere and the reduction conditions are
also documented in the same paper.

The Pt on alumina catalyst was prepared by impregnation of platinum-chloride precursor 
followed by calcination at 525 \degc~in air \cite{kelly_cjc_2019}. The powder sample was ground and pressed 
into a metal sample holder for the transmission measurements at the MRCAT \cite{segre_aip_2000} at 
Advanced Photon Source. The x-ray beam size on the sample was approximately 1 mm by 
1 mm as defined by slits. The double crystal Si(111) monochromator was scanned to 
select the x-ray energy and a Rh-coated mirror was used to remove x-rays with 
higher harmonic energies. The X-ray energy was calibrated by using the Pt 
absorption edge of Pt foil.  The custom designed in-situ cell is described 
elsewhere \cite{bare_rsi_2006}. The sample was heated from room temperature to 500\degc~in 
100\% H$_2$, resulting in continuous reduction of the Pt oxidation state. 
In-situ Pt XANES measurements were recorded in transmission during this process.

\subsection{Algorithm deployment and data collection of the NMC111 sample}

An NMC111 cathode laminate was prepared as \SI{90}{\percent} (w/w) \ce{LiNi_{1/3}Mn_{1/3}Co_{1/3}O2} (BASF TODA Battery Materials LLC NM-3101), \SI{5}{\percent} (w/w) Timcal C-45 carbon, and \SI{5}{\percent} (w/w) Solvay 5130 PVDF Binder. The slurry was coated onto a \SI{20}{\micro\meter} Al foil current collector for a combined thickness of \SI{61}{\micro\meter}. \SI{12.7}{\milli\meter} diameter electrodes were punched from the dried laminate for cycling.
For
in-situ measurements, the electrode was placed in an AMPIX cell \cite{borkiewicz_jac_2012} with a Li metal anode and \SI{1.0}{\mole\per\liter} \ce{LiPF6} in ethylene carbonate/dimethyl carbonate 50/50 (v/v) electrolyte. 

The data collection was done at the 25-ID-C beamline of the Advanced Photon Source,
where a KB mirror pair is used for focusing x-ray on the sample and a Si(111)
monochromator is used for energy selection. The monochromator is controlled through
Bluesky \cite{allan_synchrotron_rad_news_2019}, a scriptable experiment control
library in Python. Wrapping our adaptive sampling algorithm in a Bluesky plan enabled receiving measured intensities from, and sending
suggestions of energies to measure to the Bluesky run engine, thereby realizing automated
control of beamline instruments by the algorithm. 

During the experiment, the battery cell was charged then discharged galvanostatically at \SI{0.205}{\milli\ampere} (\SI{0.1}{\per\hour}) between \SI{3.5}{\volt} and \SI{4.5}{\volt} using a Maccor 4300. 8 XANES spectra were collected sequentially with
equal time interval during the process under the guidance of
our algorithm. Additionally, 2 reference spectra were measured on
the uncharged (pristine) and fully charged (\SI{4.5}{\volt}) samples.
The reference spectra were measured over \SI{\pm100}{\electronvolt}
relative to the K absorption edge of Ni (\SI{8333}{\electronvolt}), or \SIrange{8233}{8433}{\electronvolt},
with a uniform step size of \SI{0.5}{\electronvolt}.
On the other hand, adaptive sampling was done on a larger range of 
\SIrange{-100}{150}{\electronvolt} relative to the Ni edge. 

\section*{Author contributions}

MD, CS, SK and MC proposed the research and discussed through the course of the method's development. MD developed the code that implements the method, conducted computational experiments, and analyzed the data. CS contributed the LTO and YBCO data. SK contributed the Pt data. MW prepared the NMC111 sample and conducted beamline experiment with MD. 

\section*{Data availability}

The data that supports the findings of this study is available from the corresponding author upon reasonable request.
Data and code will be published with the paper once permission has been received from our institution upon the acceptance of the paper. 

\section*{Code availability}

The code that implements the proposed algorithm is available from the corresponding author upon reasonable request.
Data and code will be published with the paper once permission has been received from our institution upon the acceptance of the paper. 

\section*{Acknowledgement}

This research used resources of the Advanced Photon Source, an Office of Science User Facility operated for the U.S. Department of Energy (DOE) Office of Science by Argonne National Laboratory, and was supported by the U.S. DOE under Contract No. DE-AC02-06CH11357, and the Canadian Light Source and its funding partners. Data for demonstrations were collected at beamline 25-ID, 20-BM, and the 10-ID (MRCAT) of the Advanced Photon Source. The NMC-111 electrodes were supplied by the U.S. Department of Energy’s (DOE) CAMP (Cell Analysis, Modeling and Prototyping) Facility, Argonne National Laboratory. The CAMP Facility is fully supported by the DOE Vehicle Technologies Office (VTO).

\bibliographystyle{unsrt}
\bibliography{mybib}

\end{document}


\maketitle

\begin{itemize}
    \item[$^a$] Advanced Photon Source, Argonne National Laboratory, Lemont, Illinois 60439, USA
    \item[$^*$] mingdu@anl.gov; skelly@anl.gov; mcherukara@anl.gov
    \item[$^\dagger$] These authors contributed equally.
\end{itemize}

\begin{figure}[h]
    \centering
    \includegraphics[width=1\linewidth]{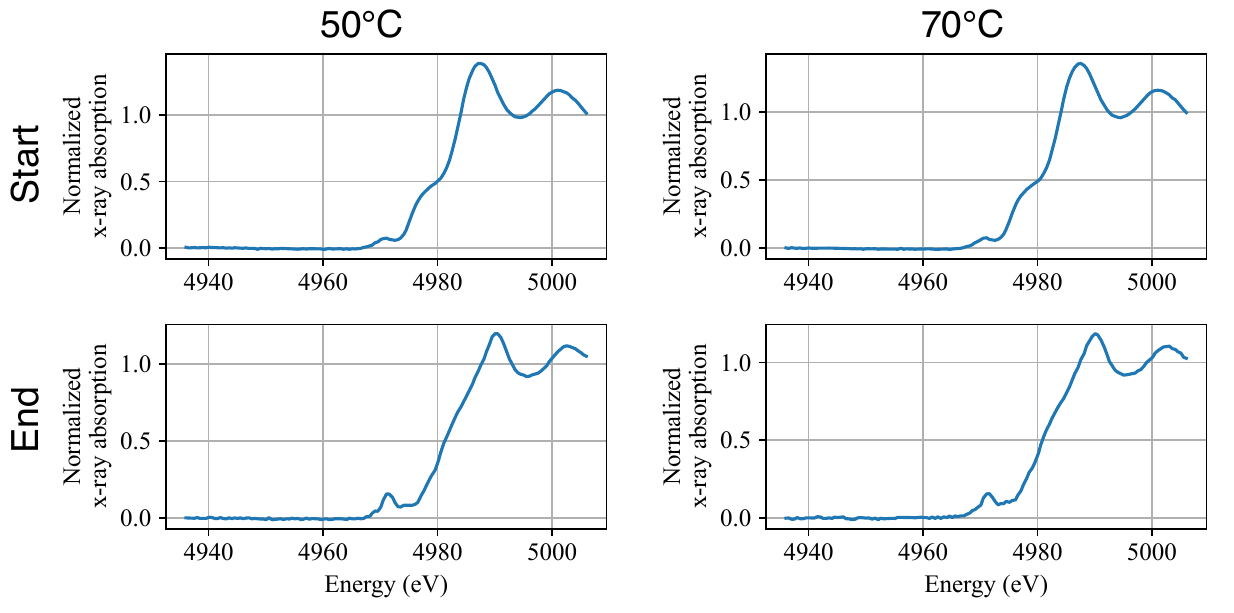}
    \caption{Normalized ground truth XANES spectra of LTO at the start and end of phase 
    transition at 50\degc~and 70\degc.}
    \label{fig:lto_normalized_spectra}
\end{figure}

\begin{figure}
    \centering
    \includegraphics[width=1\linewidth]{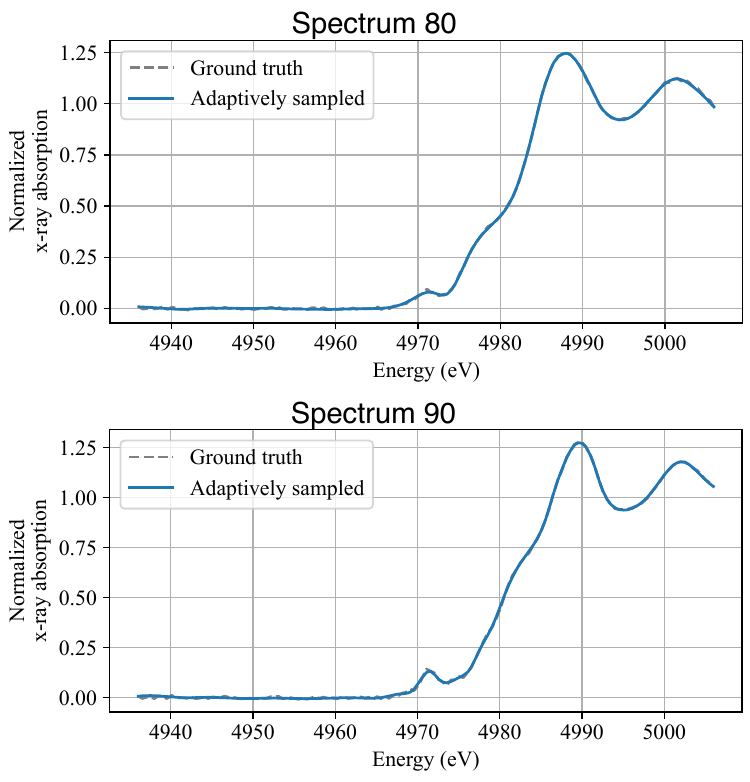}
    \caption{XANES spectra of LTO estimated with adaptively sampled data 
    at indices 80 and 90, which are immediately before and
    after the phase transition. Spectrum 90 differs from 80 mainly in the more pronounced
    pre-edge peak. Data plotted are normalized and flattened. }
    \label{fig:lto_spectra_before_after_phase_transition}
\end{figure}

\begin{figure}
    \centering
    \includegraphics[width=1\linewidth]{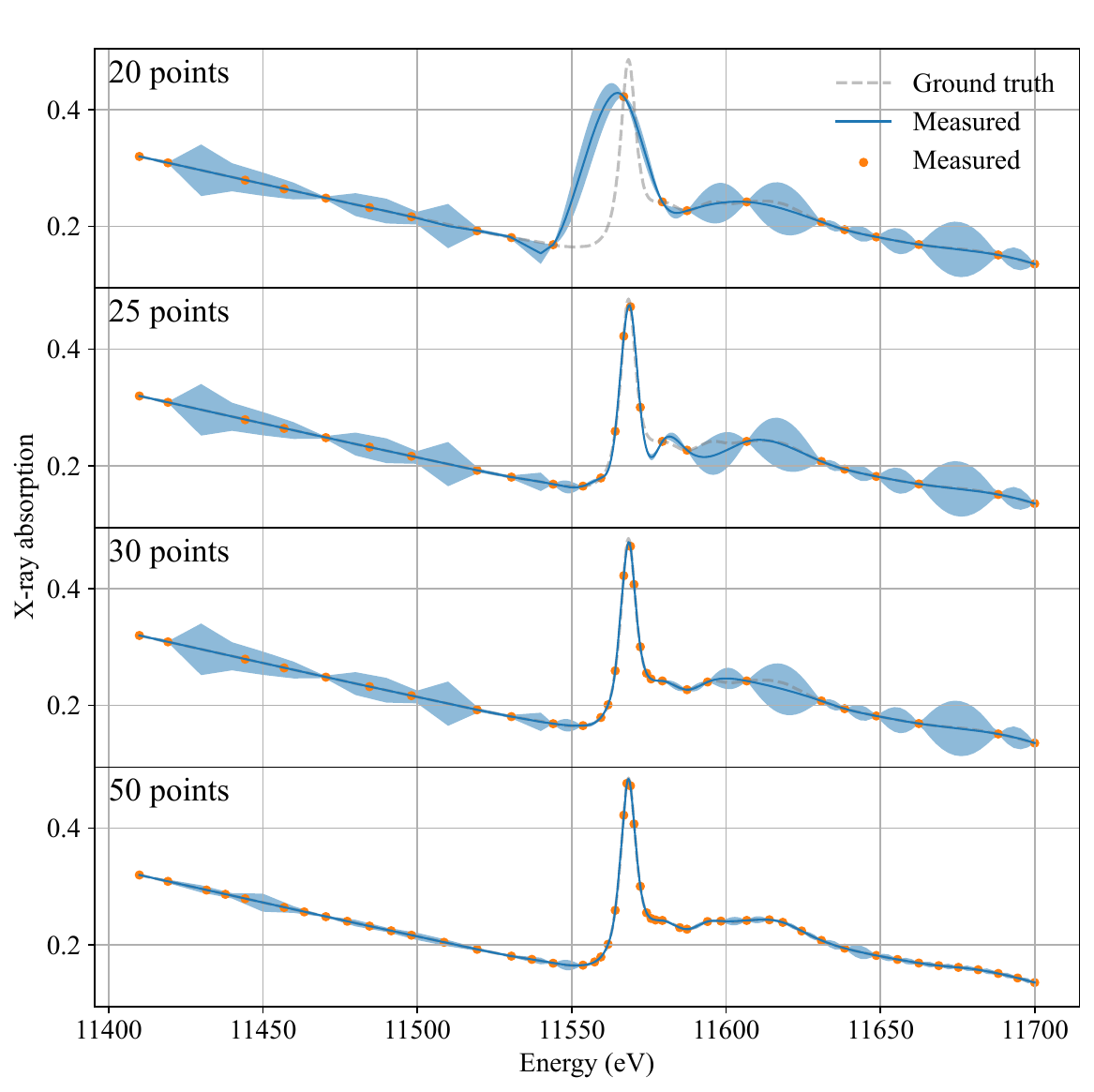}
    \caption{Intermediate reconstructed spectra, posterior standard deviation, 
    measured data points, and true spectrum for the Pt data. The posterior
    standard deviation at each point is represented by half of the vertical length of
    the shaded area. All data shown are before normalization.}
    \label{fig:pt_intermediate}
\end{figure}

\begin{figure}
    \centering
    \includegraphics[width=1\linewidth]{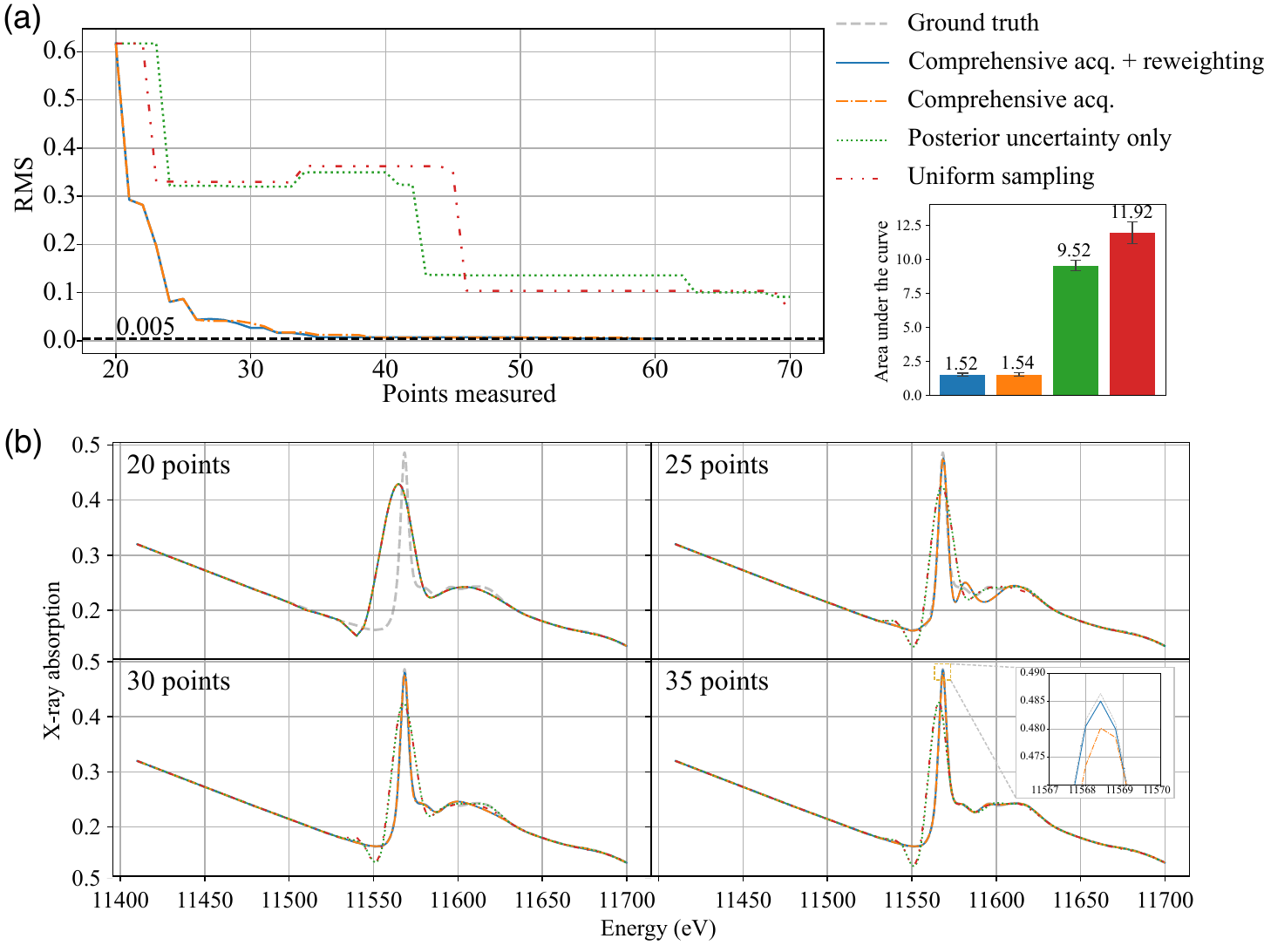}
    \caption{Results of single-spectrum sampling of the Pt sample.
    (a) Convergence of unnormalized RMS error with our comprehensive acquisition function
    and acquisition reweighting, and its comparison with cases (\emph{i}) with 
    comprehensive acquisition but without reweighting,
    (\emph{ii}) with posterior uncertainty-only acquisition function, and (\emph{iii}) with uniform sampling.
    The bar chart on the side shows the areas under the curve (AUCs) before the 55th measurement 
    for each case with
    the same color coding to the RMS error curves. The heights
    of the bars and the numbers indicate the averages of the AUCs over 5 repeated runs with different initial points, and the error bars
    represent their standard deviations.
    (b) Intermediate reconstructed spectra obtained with the 4 methods over the 15 points sampled
    after the initial measurements. Inset in the panel of 35 points zooms in around the white line peak to show the superior sampling of the white line of the method with acquisition reweighting.}
    \label{fig:pt_comparison}
\end{figure}

%% file: government_license.tex
\textbf{GOVERNMENT LICENSE}

The submitted manuscript has been created by UChicago Argonne, LLC, Operator of Argonne
National Laboratory (``Argonne''). Argonne, a U.S. Department of Energy Office of Science laboratory, is operated under Contract No. DE-AC02-06CH11357. The U.S. Government retains for
itself, and others acting on its behalf, a paid-up nonexclusive, irrevocable worldwide license in
said article to reproduce, prepare derivative works, distribute copies to the public, and perform
publicly and display publicly, by or on behalf of the Government. The Department of Energy will
provide public access to these results of federally sponsored research in accordance with the DOE
Public Access Plan. http://energy.gov/downloads/doe-public-access-plan.